\documentclass[12pt,preprint]{aastex}
\usepackage{color}

\newcommand{\kms}{km~s$^{-1}$}
\newcommand{\lsim}{{\, \lower2truept\hbox{
${< \atop\hbox{\raise4truept\hbox{$\sim$}}}$}\,}}
\newcommand{\gsim}{{\, \lower2truept\hbox{
${> \atop\hbox{\raise4truept\hbox{$\sim$}}}$}\,}}
\newcommand{\feiiq}{\rm Fe{\sc ii }$\lambda$4570\/}

\def\lledd{$L_{\rm bol}/L_{\rm Edd}$}
\def\ne{$n_{\rm e}$\/}

\def\rfe{$R_{\rm FeII}$}

\def\ltsima{$\; \buildrel < \over \sim \;$}
\def\simlt{\lower.5ex\hbox{\ltsima}}            
\def\gtsima{$\; \buildrel > \over \sim \;$}

\def\simgt{\lower.5ex\hbox{\gtsima}}            

\def\civ{{\sc{Civ}}$\lambda$1549\/}
\def\civnc{{\sc{Civ}}$\lambda$1549$_{\rm NC}$\/}
\def\civbc{{\sc{Civ}}$\lambda$1549$_{\rm BC}$\/}

\def\cm3{cm$^{-3}$\/}
\def\hb{{\sc{H}}$\beta$\/}

\def\hbbc{{\sc{H}}$\beta_{\rm BC}$\/}

\def\hbnc{{\sc{H}}$\beta_{\rm NC}$\/}
\def\hgnc{{\sc{H}}$\gamma_{\rm NC}$\/}
\def\hanc{{\sc{H}}$\alpha_{\rm NC}$\/}

\def\niv{{\sc{Niv]}}$\lambda$1486\/}
\def\ciii{{\sc{Ciii]}}$\lambda$1909\/}
\def\oiiiopt{{\sc{[Oiii]}}$\lambda\lambda$4959,5007\/}
\def\o4363{{\sc{[Oiii]}}$\lambda$4363\/}
\def\oiiiuv{{\sc{Oiii]}}$\lambda$1663\/}
\def\siiii{{Si}{\sc iii}]$\lambda$1892\/}
\def\heiiuv{{\sc{Heii}}$\lambda$1640}

\def\feiiuv{{{Fe\sc{ii}}}$_{\rm UV}$\/}

\def\feii{{Fe\sc{ii}}\/}

\def\dvr{{$\Delta v_{\mathrm r}$}}
\def\fevii6087{{\sc [Fe vii]}$\lambda$6087\/}
\def\oiii{{\sc [Oiii]}$\lambda$5007}

\def\kms{km~s$^{-1}$}
\def\ergss{ergs s$^{-1}$\/}
\def\mbh{$M_{\rm BH}$\/}
\def\gs{$\Gamma_{\rm soft}$\/}

\def\hi{H{\sc i}\/}
\def\rk{{$R{\rm _K}$}\/}

\def\ab{$A_{\mathrm B}$\/}
\begin{document}

\title{\civ\ as an Eigenvector 1 Parameter for Active Galactic Nuclei}
\slugcomment{ } \shorttitle{\civ\ in AGN} \shortauthors{Sulentic
et al. }
\author{
Jack W. Sulentic\altaffilmark{1}, Rumen Bachev\altaffilmark{1},
Paola Marziani\altaffilmark{2}, \\ C. Alenka
Negrete\altaffilmark{3}, Deborah Dultzin\altaffilmark{3} }

\altaffiltext{1}{Department of Physics and Astronomy, University
of Alabama, Tuscaloosa, AL 35487, USA; giacomo@merlot.astr.ua.edu
}

\altaffiltext{2}{INAF, Osservatorio Astronomico di Padova, Vicolo
dell'Osservatorio 5,   I-35122 Padova, Italy;
paola.marziani@oapd.inaf.it}

\altaffiltext{3}{Instituto de Astronom\'\i a, Universidad Nacional
Autonoma de M\'exico (UNAM), Apdo. Postal 70-264, 04510 Mexico
D.F., Mexico; deborah@astroscu.unam.mx}

\begin{abstract}

We have been exploring a spectroscopic unification for all known
types of broad line emitting AGN. The 4D Eigenvector 1 (4DE1)
parameter space shows promise as a unification capable of
organizing quasar diversity on a sequence primarily governed by
Eddington ratio.  This paper considers the role of \civ\ measures
with special emphasis on the \civ\ line shift as a principal 4DE1
diagnostic.  We use HST archival spectra for 130 sources with
$S/N$\ high enough to permit reliable \civ\ broad component
measures. We find a \civbc\ profile blueshift that is strongly
concentrated among (largely radio-quiet: RQ) sources with
FWHM(\hbbc) $\la$ 4000 \kms\ (which we call Population A). Narrow
line Seyfert 1 (NLSy1, with  FWHM \hb\ $\leq$ 2000 \kms) sources
belong to this population but do not emerge as a distinct class.
The systematic blueshift, widely interpreted as arising in a disk
wind/outflow, is not observed in broader line AGN (including most
radio-loud (RL) sources but also $\sim$ 25 \%\ of RQ) which we
call Population B. We find new correlations between FWHM(\civbc)
and \civ\ line shift as well as the equivalent width of \civ. They
are seen only in Pop. A sources.  Broader-lined sources show
random scatter. \civ\ measures enhance the apparent dichotomy
between sources with FWHM(\hbbc){ \em less and greater than }\
4000 \kms\ \citep{sulenticetal00a} suggesting that it has more
significance in the context of Broad Line Region structure than
the more commonly discussed  RL vs. RQ dichotomy. Black hole
masses computed from FWHM \civbc\ for about 80 AGN indicate that
the \civ\ width is a poor virial estimator. Comparison of mass
estimates derived from \hbbc\ and \civ\ reveals that the latter
show different and nonlinear offsets for population A and B
sources. A significant number of sources also show narrow line
\civ\ emission that must be removed before \civbc\ measures can be
made and interpreted effectively. We present a recipe for \civ\
narrow component extraction.

\end{abstract}


 \keywords{quasars: general, emission lines; line:
profiles}


\section{Introduction}

The search for a parameter space that might provide spectroscopic
unification for all classes of broad line emitting AGN motivated
the ``4D Eigenvector 1" (4DE1) concept
\citep{sulenticetal00a,sulenticetal00b}. Such a correlation space
might serve as an equivalent to the stellar H-R Diagram. Domain
space occupation differences and parameter correlations might then
provide the empirical clues from which underlying physics could be
inferred. At the very least it can be used to highlight important
differences between sources that can also drive our physical
understanding of the geometry, kinematics and physics of the broad
line emitting region (BLR). From the outset it was expected that a
parameter space for AGN would require more than two dimensions
because source orientation and ``physics'' (e.g., black hole mass
\mbh\ and Eddington ratio) drive AGN parameter values and
correlations. A suitably chosen $n$-dimensional space should help
to remove the degeneracy between these two drivers.

4DE1 has roots in the PCA analysis of the Bright Quasar Sample
\citep[87 sources; ][]{borosongreen92} as well as in correlations
that emerged from ROSAT \citep[e.g. ][]{wangetal96}. 4DE1 as we
define it involves BG92 measures: (1) full width half maximum of
broad \hb\ (FWHM \hb) and (2) equivalent width ratio of optical
\feii\ and broad \hb: \rfe = W(\feiiq)/W(\hbbc). We added a
\citet{wangetal96}-defined measure involving (3) the soft X-ray
photon index (\gs) and a  measure of (4) \civ\ broad line profile
velocity displacement at half maximum ($c(\case{1}{2})$) to arrive
at our 4DE1 space. Other points of departure from BG92 involve our
comparison of RQ and RL sources as well as subordination of BG92
\oiii\ measures \citep[although see
][]{zamanovetal02,marzianietal03a}.  Finally we divide sources
into two AGN populations using a simple division at FWHM \hbbc\ $
=$4000 \kms\  with sources narrower and broader than this value
designated Pop. A and B respectively. It was motivated by the
observation that almost all RL sources show FWHM\hbbc $\ga$ 4000
\kms\ \citep{sulenticetal00b}.  This division appears to be more
effective than the more traditional divisions into: (1) RQ-RL
sources  as well as (2) NLSy1 sources defined with
FWHM(\hbbc)$\la$ 2000 \kms\ and broader line sources above this
value. Results presented in this paper strongly support the Pop.
A-B distinction. Exploration of possible physical drivers of
source occupation/correlation in 4DE1
\citep{marzianietal01,marzianietal03b,boroson02}  suggest that it
is primarily driven by the luminosity to black hole mass (\mbh)
ratio which is proportional to the Eddington ratio (\lledd) with
Pop. A sources  being high accreting/low \mbh\ AGN, while Pop. B
being low accreting/large \mbh\ AGN.

Past 4DE1 studies focused on the optical 4DE1 plane (FWHM \hb\ vs.
\rfe) at low redshift because more high S/N optical spectra exist
than UV or X-ray measures. Complementary high-$z$\ measures of the
\hb\ region at IR wavelengths are ongoing
\citep{sulenticetal04,sulenticetal06a}. This paper focuses on an
expanded sample of \civ\ measures and explores their utility as
4DE1 parameters. The work is supplemental to a recent paper
\citep{bachevetal04} that discusses data processing and analysis
of 123 \civ\ spectra from the HST archive. The new \civ\ sample is
almost twice the size of the one discussed in the defining 4DE1
paper \citep{sulenticetal00a}. We present (\S \ref{civ}) new 4DE1
correlation diagrams involving measures of the \civ\ line shift
and then look (\S \ref{e1}) at the implications of \civ-defined
source occupation for BLR structure and for the hypothesized AGN
Populations (A and B; \S \ref{pop}). Section \ref{nc} discusses
the reality of a significant narrow line \civ\ component in many
sources and compares our \civ\ measures with other recent studies
utilizing the same spectra. Section \ref{mbh} considers the
implications of our \civ\ results on the use of FWHM \civ\ to
estimate black holes masses.

\section{\civ\ Line Measures and Correlations}
\label{civ}

\subsection{Sample Definition and Data Analysis}

We searched the HST archive\footnote{Datasets covering the \civ\
sources listed in Table 1 can be all identified and retrieved from
the WWW site at URL \url{http://archive.stsci.edu/hst} and are not
reported here. A list with the actual datasets employed is
available from the authors at URL
\url{http://web.oapd.inaf.it/marziani}.} and found useable \civ\
spectra for 130 out 141 low-redshift sources. Excluded sources are
mostly \civ\ BAL quasars where reliable measures of the \civ\
emission profile are difficult. OI 363 was not included because of
low S/N. IRAS 13218+0552 (J132419.9+053705) was excluded because
it shows no broad lines that would warrant a Type 1 AGN
designation. We assume that our sample is large enough to
reasonably represent the broad emission line properties of low
$z$\ AGN. It is likely to be the only UV dataset of reasonable
quality quasar spectra in the foreseeable future. The sample
should be particularly valuable for RQ vs. RL comparisons because
the two populations are almost equally represented while in a
complete sample only $\approx$ 10 \% are found to be RL
\citep{jiangetal06,cirasuoloetal03,sulenticetal03}.  A PG quasar
subsample was identified and includes 43 sources with 26\% RL
reflecting the overrepresentation of RL sources in the HST
archive.

The uncertainty due to instrumental errors in wavelength
calibration are estimated to be $\approx 200$ \kms\
\citep{marzianietal96}. In order to reduce wavelength calibration
errors HST spectra were ``re-aligned" using expected
rest-wavelengths of strong low-ionization, Galactic
absorption-lines including \ion{Mg}{2} $\lambda$2796.35,
\ion{Mg}{2} $\lambda$2803.53, \ion{Fe}{2} $\lambda$2600.17,
\ion{Fe}{2} $\lambda$2586.65, \ion{Fe}{2} $\lambda$2382.77,
\ion{Fe}{2} $\lambda$2374.46, \ion{Fe}{2} $\lambda$2344.21,
\ion{Al}{2} $\lambda$1670.79, \ion{Si}{2} $\lambda$1526.71,
\ion{C}{2} $\lambda$1334.53, \ion{Si}{2} $\lambda$1260.42
\citep{savageetal00}. In case only one or two Galactic lines were
available in the spectra, any shift between expected Galactic line
wavelength and the wavelength measured on the spectra was double
checked to avoid spurious results due to low S/N. Suitable
Galactic lines were found for 110 sources in our sample with three
or more lines available for 71 sources.  The average rms of the
residuals between measured line wavelengths after re-alignment and
tabulated wavelengths is $< rms
> \approx 40 $ \kms. This provides an estimate of the wavelength
calibration uncertainty (at 1$\sigma$\ confidence level) for the
re-aligned spectra.

The broad component of \civ\ (\civbc) was extracted after
correction for contaminating lines (\niv, and especially \heiiuv
and \oiiiuv) and subtraction of \feiiuv\ emission  (details of
data reduction are given in \citet{bachevetal04} and
\citet{marzianietal96}). The continuum underlying \civ\ was
estimated from nearby regions that are free of strong emission
lines (between the $\lambda$1400 blend and \niv\ on the blue side
as well as 1700 -- 1800 \AA\ on the red side).  A narrow component
(\civnc) was subtracted from the profile when warranted. There is
still disagreement about the existence, frequency of occurrence
and strength of any \civnc. We discuss the evidence for NLR \civ\
and describe our \civnc\ subtraction procedure in \S \ref{nc}.

\subsection{Immediate Results}
\label{immres}

Fig. \ref{fig:civbc} shows individual cleaned \civbc\ profiles fit
with high-order spline functions to minimize effects of noise and
to preserve the complexity of the shape \citep[following
][]{marzianietal96,marzianietal03a}. The spline fit is shown as a
thick line in Fig. \ref{fig:civbc} while identified narrow
components (that were subtracted in this analysis) are seen above
the spline.

Table 1 gives an identification list of all sources shown in
Figure 1 along with 4DE1 optical and X-ray parameters. Table 1
includes: Column 1 -- IAU code identification; Col. 2 - a common
name for the source; Col. 3 -- available source redshift with
number of significant figures indicating accuracy of the
determination;  Col. 4 -- redshift reference. Col. 5 -- an
asterisk indicates that the sources belongs to the
\citet{borosongreen92} PG sample, a ``B" indicates that the source
is a ``blue outlier" \citep{zamanovetal02}. Col. 6 -- Galactic
absorption (\ab, in magnitudes), Col. 7 -- available measures of
FWHM for \hb\ broad component  (FWHM(\hbbc), units \kms) taken
from \citet{marzianietal03b}, measures of SDSS spectra or, as a
last resort, literature spectra; Col. 8 --  measures of the ratio
\rfe\ from same sources as Col. 7;  Col. 9 -- decimal logarithm of
the specific flux at 4400 \AA\ over the flux at 6 cm. A source is
assumed radio-loud  if $\log$ \rk $\ge$ 1.8;  Col. 10 -- a measure
of the soft X-ray excess (photon index \gs), from
\citep{sulenticetal00a,sulenticetal00b} and from various
literature sources.

The reported optical redshifts come from measures of
low-ionization optical emission lines (LILs), typically \hbnc,
\hgnc, and \hanc\, supplemented by values derived from \oiiiopt\
if the source is not a blue outlier \citep[see
][]{marzianietal03a,zamanovetal02}. In these cases, the agreement
between LIL and high-ionization lines (HILs) is reasonable within
the accuracy limits of the present study. We remind that ``blue
outliers" i.e., sources with large \oiiiopt\ blueshift relative to
optical LILs, tend to be extreme Pop.  A sources with very weak
\oiiiopt, and are relatively rare. The recipe described in
\citet{marzianietal03a} is applied for all sources with references
indicated as ESO, SPM, M03, M96, SDSS, G99. All other sources have
redshift measured on the basis of the optical lines. None of the
remaining sources are likely to be blue outliers on the basis of
published spectra so redshift computed using optical lines should
be a reliable estimate even if \oiiiopt\ lines were used.

Table 2 presents our \civ\ parameter measures with format as
follows: Column 1 -- IAU code; Col. 2 -- specific continuum flux
at 1550$\rm{\AA}$ (units erg s$^{-1}$ $\rm{\AA}^{-1}$
cm$^{-2}$$\times$10$^{14}$); Col. 3 -- flux in the \civnc\ (units
erg s$^{-1}$ cm$^{-2}$$\times$10$^{13}$); Col. 4 -- Peak \civnc\
radial velocity, in \kms;  Col. 5 -- flux in the \civbc\ (units as
in Col. 3); Cols. 6, 7, 8 -- centroid profile shift at
$\case{1}{4}$ maximum ($c$($\case{1}{4}$)) followed by the
estimated uncertainties on the blue and red wings of the profile
(units \kms); Cols. 9, 10,11 -- same at half-maximum
($c(\case{1}{2}$), which is an adopted 4DE1 parameter); Cols. 12,
13 -- centroid at $\case{3}{4}$\ maximum ($c$($\case{3}{4}$)) with
symmetric uncertainty ; Cols. 14, 15 -- centroid at the 90\%\
intensity level of the \civ\ broad line (c(0.9)), with symmetric
uncertainty; Cols. 16, 17 -- FWHM(\civbc) and estimated
uncertainty (units \kms); Cols. 18,19, 20 -- \civbc\ asymmetry
index with estimated uncertainties on the blue and red profile
wings; Cols. 21,22 -- \civbc\ kurtosis measure and estimated
uncertainty.

No \civnc\ measures are given in Table \ref{tab:civ} if the
profile if affected by  partial (a) or strong (A) absorption. \ In
sources labeled a in Tab. \ref{tab:civ} residual \civnc\ is
sometimes visible but the NC width and flux cannot be recovered.
NC shifts and fluxes are accurate  (within $\pm$ 40\%\ at a
2-$\sigma$\ confidence level) only if \civnc\ emission shows an
intensity at least 0.05 \civbc. Note that our adopted \civnc\
component is often not ``\oiiiopt-like". It is often significantly
broader and stronger than would be subtracted if we used \oiii\ as
a template for the \civ\ doublet. See  \S \ref{nc} for both
empirical and theoretical justifications for our procedure.

Measured centroids at different fractional intensities were defined as
follows:

\begin{equation}
\label{ } c(\case{i}{4}) = \frac{\lambda_{\mathrm B}  +
\lambda_{\mathrm R} - 2 \lambda_0}{ 2  \lambda_0} ~c, \forall i =
0, \ldots 4,
\end{equation}
where $c$\ is the speed of light. Values $c(\case{i}{4})$\  for $i
= 0 $\ are not listed in Tab. \ref{tab:civ} due to the difficulty
of assessing $\lambda_{\mathrm B}$\ and $\lambda_{\mathrm R}$\  at
zero intensity. We give $c(\case{9}{10})$ instead of peak radial
velocity. This  has been shown to be a good surrogate and less
dependent on \civnc\ subtraction as well as line profile
irregularities \citep{marzianietal03a}. The asymmetry index is
defined as follows:
\begin{equation}
\label{ } A.I. = \frac{\lambda_{\mathrm B}(\case{1}{4})  -
\lambda_{\mathrm R}(\case{1}{4})  - 2 \lambda_{\mathrm P}}{
\lambda_{\mathrm P}}
\end{equation}
where for $ \lambda_{\mathrm P}$ we use $c(\case{9}{10})/c$.
The kurtosis index is defined as:
\begin{equation}
\label{ }
kurt = \frac{\lambda_{\mathrm R}(\case{3}{4})  - \lambda_{\mathrm B}(\case{3}{4})}{\lambda_{\mathrm R}(\case{1}{4})  - \lambda_{\mathrm B}(\case{1}{4})}
\end{equation}
\citep[cf. ][]{marzianietal96}.

Uncertainties reported in Tab. \ref{tab:civ}  were estimated by
measuring the wavelengths $\lambda_{\mathrm R}$\ and
$\lambda_{\mathrm B}$\ at $\pm$ 5\% fractional intensity and then
quadratically propagating the errors in the relationships reported
above.  All uncertainties reported in Table \ref{tab:civ}
represent a 2$\sigma$\ confidence level. Uncertainties in
estimating the rest frame velocity, relative to which the
centroids are computed, can be as large as 300 \kms\ or as small
as $\sim$ 30 \kms\ (at 1$\sigma$\ confidence level) depending on
the availability of moderate resolution spectra (SDSS is, or will
be, improving the situation for about 50\% of the sample). The
error in estimating the local rest frame $\Delta z \approx 0.00014
\pm 0.0006$ was derived from the distribution of differences
between $z$\ values used in this work and those given in NED.
Combining the typical uncertainty on systemic velocity, on UV
wavelength calibration, and the average of the measurement
uncertainty reported in Table \ref{tab:civ}, the typical
uncertainty (at a 2-$\sigma$\ confidence level) are $\approx$ 230
\kms and $\approx 170 $ \kms\ for $c(\case{1}{2})$ and
$c(\case{3}{4})$.

\subsection{\civ\ Line Parameters in the RQ-RL Context}
\label{e1}

Figure \ref{fig:e1} shows source occupation in 4DE1 planes
involving the $c(\case{1}{2})$\ parameter \citep[as defined in
][]{sulenticetal00b}. $c(\case{1}{2})$\ was chosen from among
possible \civ\ profile measures (FWHM, $c(\case{1}{2})$\ and
equivalent width) because: 1) it is not obviously luminosity
dependent, 2) it showed the largest intrinsic dispersion and 3) it
showed possible correlations with the other principal 4DE1
parameters.  As a luminosity normalized measure W(\civbc) is ruled
out even if the well-known ''Baldwin effect'' now appears to be
driven by dependance on the Eddington ratio
\citep{bachevetal04,baskinlaor04}. This does not mean that we
regard it as an unimportant measure  but only that we reject it as
one of the principal 4DE1 parameters. A surrogate measure might
involve a direct measure of \civbc\ line flux but the parameter
dispersion of that measure is less than for  $c(\case{1}{2})$. The
same is true for FWHM(\civbc) which also shows less dispersion
than FWHM(\hbbc). Line broadening may be due to both rotational
and non-rotational velocity components especially if a disk + wind
model is applicable to our sources.  $c(\case{1}{2})$\, on the
contrary,  is most likely related to the amplitude of any
non-virial motions in the BLR. It is this parameter that adds a
new element that can be argued to be {\em physically orthogonal}
to previously defined E1 parameters: FWHM(\hbbc) estimates the
virial broadening in the LIL-emitting part of the BLR; \rfe\
measures the ionization conditions, while \gs\ provides a
measurement of the continuum shape.

RQ and RL sources are indicated by circles and squares respectively in
Fig. \ref{fig:e1} \citep[sources with radio/optical flux ratio log
R$_K$ $\ga$ 1.8 are considered RL: ][]{sulenticetal03}. The large
number of squares reflects the over-representation of RL sources in our
sample.  Figure \ref{fig:e1}a shows that sources with \civ\ profile
blueshifts strongly favor RQ AGN with FWHM \hbbc$\la$ 4000 \kms. RL
sources show a large scatter of both red and blue \civ\ shifts. Figures
\ref{fig:e1}bc show that sources with \civ\ blueshift especially favor
RQ sources with large \rfe\ (strong optical \feii\ emission) and
\gs\ (a soft X-ray excess) measures respectively.  RL sources are much
more strongly concentrated in the latter two 4DE1 planes.

Table \ref{tab:ave}  gives mean parameter values (sample standard
deviations in parenthesis) for total sample, RQ, RL and our
previously defined Pop. A-B subsamples that will be considered in
the next section. Values are given for: Col. 2 -- number of
sources; Col. 3 -- equivalent width measure of the \civbc\ line;
Col. 4 -- $c(\case{1}{2})$\ of \civbc; Col. 5 -- FWHM(\civbc);
Col. 6 -- FWHM(\hbbc); Col. 7 -- \rfe\ and Col. 8 -- \gs. Columns
4,6,7 and 8 represent the principal parameters in 4DE1.  We find
that RL sources show broader \hbbc\ and \civbc\ profiles than RQ
AGN while RQ sources show stronger \rfe, \gs\ and $c(\case{1}{2})$
(blueshift) than the RL sample. FWHM \civbc\ is on average (17\%)
broader than FWHM \hbbc\ for RQ sources while FWHM \hbbc\ is
(16\%) narrower than FWHM \civbc\ for RL sources.  These
differences relate to one of the most significant results of our
earlier work where a restricted optical domain space occupation
was found for RL sources.  Figure \ref{fig:e1} also shows this
restricted occupation as a strong concentration of RL sources in a
small region of the $c(\case{1}{2})$ vs. \rfe\ and \gs\ planes. RL
sources are rarely found with 4DE1 parameter values: FWHM(\hbbc)
$\la $4000 \kms, \rfe $\ga$ 0.3, \gs $\ga$2.5 and
$c(\case{1}{2})$\ $\la $0 \kms. The expanded \civ\ sample confirms
and strengthens this result which likely indicates a fundamental
difference in BLR structure, kinematics and/or physics between RL
and RQ populations \citep[see ][ for discussion in the context of
a RQ-RL dichotomy]{sulenticetal03}.

Spearman rank correlation coefficients and associated probabilities are
given in Table \ref{tab:cc} for \civ\ equivalent width, FWHM and
centroid measures versus the three other principal 4DE1 parameters. The
total-sample correlation coefficients for this sample are larger than
corresponding values given for the smaller sample of sources in
\citet{sulenticetal00b} as one might hope to see if the correlations
are in some sense real. Table \ref{tab:cc} emphasizes the spectroscopic
differences between RQ vs RL sources by showing no evidence for
correlations among 4DE1 parameters for RL sources. Real or marginal
correlations are only found among RQ sources with the strongest
correlations involving $c(\case{1}{2})$, \rfe\ and \gs. Restriction to
a BG92 overlap subsample shows no significant difference in correlation
coefficients.


\section{Evidence for Two Populations of Broad Line AGN}
\label{pop}

So far we have compared sources on the conventional basis of a RQ
vs. RL dichotomy however it is important to point out that about
25\%\ of RQ sources in our sample occupy the same 4DE1 parameter
domain as the RL AGN. If 4DE1 parameters reflect broad line
physics/kinematics then this overlap may be important. The
restricted 4DE1 parameter space occupation for RL sources
motivated us \citep{sulenticetal00b} to hypothesize the existence
of two AGN ``populations'' (A and B) defined in an optical
spectroscopic context (4DE1) rather than on the basis of radio
loudness. Following our scheme Population A sources show
FWHM(\hbbc) $\la$ 4000 \kms, strong \rfe, strong \gs\ (a soft
X-ray excess) and a $c(\case{1}{2})$\ blueshift with estimated
probability of radio loudness $P \la$ 0.01.  Population B sources
show FWHM(\hbbc) $\ga$ 4000 \kms, weak \rfe, no soft X-ray excess
or \civ\ blueshift with estimated probability of radio loudness $P
\approx $0.30.  Revisiting Fig. \ref{fig:e1}a in the population
A-B context shows that \civ\ blueshifts are strongly concentrated
among Pop. A sources with FWHM(\hbbc) $\la$ 4000 \kms. Filled and
open symbols in Figures  \ref{fig:e1} identify Pop.  A and B
sources respectively. It is important to point out that
FWHM(\hbbc) = 4000 \kms\ was chosen as a Pop. A-B boundary before
$c(\case{1}{2})$\ was selected as an 4DE1 parameter. Pop. A-B is
more effective than the RQ-RL distinction for highlighting
spectroscopic differences.

Pop. B sources show a scatter of line shifts within
$c(\case{1}{2})$\ = $\pm$ 2000 \kms\ (Table \ref{tab:civ}) with
mean value in Table \ref{tab:ave} consistent with zero shift. A
large part of the Pop. B scatter may be  associated with \civ\
measurement uncertainties (the 3$\sigma$ shift uncertainty is
$\approx$ 400 \kms).  Fig. \ref{fig:e1} show that Pop. A sources
have a wider parameter dispersion than Pop. B sources. The
majority of Pop. B sources are so concentrated that one can assign
unique (within measurement errors) values of \rfe $\sim$ 0.15
$\pm$ 0.15 and  \gs $ \approx 2.1 \pm$ 0.5. to the entire
population. These two values along with $c(\case{1}{2})$ = --70
(consistent with 0) $\pm$ 1000 \kms\ represent the 4DE1
coordinates with highest probability of radio-loudness.

The strong parameter concentration of Pop. B sources relative to
the Pop. A RQ majority reenforces the interpretation
\citep{sulenticetal03} that RL quasars represent a distinct AGN
population and perhaps the endpoint of quasar activity in sources
with largest \mbh\ and lowest \lledd. The obvious question then
involves the relationship between Pop. B RQ sources and Pop. B RL
AGN.  RL and RQ Pop. B sources show strong similarity in most
properties but $c(\case{1}{2})$\ suggests a possible small
separation with mean \civbc\ shift values  of about --200 and
$+$70 \kms\ respectively for Pop. B RQ and RL sources. However
both values are consistent with zero shift given measurement
uncertainties. A $K-S $\ test for the two $c(\case{1}{2})$\
distributions confirms no significant difference. If \civbc\
blueshift and negative asymmetry index are the signature of a disk
wind that is driven by high \lledd\ then 60-80\% of RQ and very
few RL, sources show evidence for it. RQ  sources in our sample
have an average negative asymmetry index (--0.1), and a K-S test
confirms a significant difference with the distribution for RL
sources (whose average is +0.08). The simplest answer to the above
question then would be that for a given \mbh, RL sources lie at
the extreme low end of an \lledd\ sequence -- perhaps they are
expiring quasars. Perhaps Pop. B RQ sources with lowest values of
\civbc\ shift are the RQ expiring quasars. In that case our
population B designation has a physical significance although we
do not yet know what physical property allows/inhibits radio loud
activity.

\subsection{Population A and NLSy1s}

The distinction between Pop. A and B may be more fundamental than
RQ-RL or NLSy1-BLSy1. Fig. \ref{fig:e1} shows that \civ\
blueshifts are equally divided between sources with FWHM(\hbbc)
$\la$ 2000 \kms\ (traditional NLSy1) and sources with FWHM(\hbbc)
in the range 2000--4000 \kms\ (BLSy1). A K-S test reveals no
significant distribution difference between the two groups of
sources suggesting that the 2000 \kms\ cutoff for NLSy1 is
artificial. The same is true for comparisons involving W(\civbc),
FWHM(\civ) and \gs\ parameter distributions. Only \rfe\ seems
shows possible evidence for a difference ($D_{\mathrm KS} \approx
1.65$ with probability $P \approx 0.01$ that the two \rfe\ data
sets are not drawn from the same parent population).  Caution is
needed because the precision of the \rfe\ measure depends in both
$S/N$ and the line width \citep{marzianietal03a}. Considering the
similarity in \gs\ and \civ\ centroid shifts (which are likely
related accretion rate and disk wind properties) the Pop. A/B
distinction can be viewed as a physically motivated re-definition
of the NLSy1/BLSy1 boundary originally introduced by
\citet{osterbrockpogge85}  and subsequently adopted by 4DE1 for
different (RL) reasons.

\subsection{{Population Subdivision and Quasar Structure}}

Tables \ref{tab:ave} and \ref{tab:cc} show that the Pop. A-B
discrimination is more effective than the RQ-RL distinction for
emphasizing source differences. Table \ref{tab:ave} shows that
almost all sample mean differences between Pop. A and Pop. B are
{\it larger} than equivalent differences between RQ  and RL. Since
the entire RQ source population shows a larger parameter spread
than Pop. A RQ sources alone, it  should be {\em more} sensitive
to correlation than Pop.  A RQ alone.  Table \ref{tab:cc} confirms
that in no case does the entire RQ sample show a higher
correlation coefficient than Pop. A sources alone. In all cases
the correlation coefficient improves (or remains the same) when we
restrict the RQ sample to Pop. A RQ alone. We interpret these
results as support for our hypothesis that the Pop. A-B
distinction is {\em more fundamental} than the RQ-RL one.  So far
we have distinguished between Pop. A and B sources using FWHM
\hbbc\ alone. The mean values given in Table \ref{tab:ave} allow
us to give best estimates for the Pop. A-B boundary using the
other three principal 4DE1 parameters: \rfe $\approx$ 0.4, \gs
$\approx$ 2.60 and $c(\case{1}{2}) \approx$ 0 \kms.

1D projections of the 4DE1 space like Fig. \ref{fig:e1}abc show a main
sequence of source occupation/correlation. The Population A-B concept
reflects either a continuous variation in physical/geometric/kinematic
properties along this sequence or a true source dichotomy possibly
driven by a critical value of \lledd\ (with Pop. B RQ-RL dichotomy due
perhaps to BH spin, host galaxy properties and  role for secular
evolution in BH growth). In the former case Pop. A-B remain useful as a
vehicle for emphasizing source extrema providing a valuable challenge
to models of BLR structure/kinematics as well as changes in them due to
physics and/or source evolution
\citep{marzianietal01,marzianietal03b,boroson02}. In the past few years
we have favored the  possibility of two disjoint AGN populations on the
basis of multifold evidence:

\begin{itemize}
\item A possible gap or  paucity of sources with FWHM(\hbbc)
$\approx$ 4000 \kms, which is appreciable also in e.g., Fig. 6 of
\citet{wangetal96}, Fig. 2 of \citet{boller04},
in Fig. 1 of
\citet{sulenticetal00b}, in Fig. 3 of \citep{baskinlaor05}, most
impressively in Figure 7 (right panel) of \citep{corbor96} and see
also \citet{collinetal06}.

\item Most RL sources lie above FWHM \hbbc\ $\approx$ 4000 \kms\
while most RQ sources lie below this value. The few RL with FWHM
\hbbc\ $\la$ 4000\kms\  are likely viewed at  an orientation that
minimizes any rotational component associated with BLR motions
\citep[][ they fall there because of orientation rather than
physics]{sulenticetal03}.

\item Sources with FWHM(\hbbc) $\la 4000$ \kms\ show average
profiles well fit by a Lorentzian function while broader line
sources show profiles that are frequently red-ward asymmetric and
that require two Gaussians for a reasonable fit
\citep{sulenticetal02}.

\item Sources with FWHM(\hbbc) $\la$ 4000 \kms\ often show a soft
X-ray excess (\gs $\ga$ 2.8) while those above this limit almost
never show one \citep{boller04,sulenticetal00a}.

\item Sources with weak (usually less than 10 \AA\ equivalent
with) and blueshifted \oiii\ (the so-called blue outliers) are
found only in sources with FWHM(\hbbc) $\la$ 4000 \kms\
\citep{zamanovetal02,marzianietal03b}.

\item All sources with \civ\ $c(\case{1}{2}) \la$ --3000 \kms\
show FWHM \hbbc\ $\la$4000 \kms\ (Fig. \ref{fig:e1}a).  Most
sources with \civ\ $c(\case{1}{2}) \la$ --1000 \kms\ also lie
below the same FWHM limit. Sources with broader \hbbc\ show a
scatter of values between \civ\ $c(\case{1}{2}) \pm$ 2000\kms.
W(\civ) measures also show a strong difference (not correlation)
in mean values for source greater and less than FWHM\hbbc\  = 4000
\kms.

\item Comparison of  \civ\ and \hbbc\ profiles suggest a
discontinuity at FWHM(\hbbc)$\approx$ 4000 \kms. FWHM(\hbbc) and
FWHM(\civbc) are correlated above this value but not below
\citep[][ see also Table
\ref{tab:cc}]{marzianietal96,baskinlaor05}.

\item Fig.  \ref{fig:civ}  shows a  possible new correlation
between \civ\ FWHM and $c(\case{1}{2})$\ measures. Comparison of
Fig. \ref{fig:civ}a and \ref{fig:civ}b for Pop. B and A sources,
respectively, indicates that the correlation exists only for
sources with FWHM(\hbbc) $\la $4000 \kms\ (Pop. B sources show a
scatter diagram) . This \civ\ inter-correlation for Pop. A sources
shows a reasonably strong correlation (corr. coeff. $\approx$
0.5). The best fit relation is $c(\case{1}{2})$(\civbc) = 963 --
0.426 FWHM(\civbc) [\kms].
\end{itemize}

The correlation in Fig. \ref{fig:civ} might be expected from (and
constraining of) models that view Pop. A sources as the highest
accreting AGN that generate a disk wind
\citep{murrayetal95,bottorffetal97,progakallman04}. Previous
results may also indicate a change in BLR structure perhaps at a
critical value of \lledd\ (corresponding to FWHM(\hbbc) $\approx$
4000 \kms) with an accretion disk + outflowing high-ionization
wind required to explain Pop. A source measures
\citep{marzianietal96, marzianietal01, marzianietal03a}. Pop. B
sources do not allow us to rule out the possibility of a single
stratified emission region producing both LILs and HILs. Pop. A
and B sources differ in almost every mean property that can be
defined. Table \ref{tab:summ} summarizes both phenomenological
differences (mean values given where available) as well as some
physical differences (preceded by $\bullet$) that can be inferred
from the empiricism. Note that not all of the cited works make a
distinction between Pop. A and B.

Table \ref{tab:ave} shows that W(\civbc) differs by a factor of
$\approx$ 2 between Pop. A and B sources, with Pop. A sources
showing lower values. Since Pop. A and B do not show a significant
difference in mean source luminosity \citep{bachevetal04} we
ascribe the EW difference to a difference in \lledd which is known
to be stronger than the luminosity dependence
\citep{bachevetal04,baskinlaor04}. Pop. A and B differ
systematically in \lledd, as shown by
\citet{marzianietal03b,marzianietal06}. While not a principal 4DE1
parameter it is clear that W(\civbc) is an important measure.

\section{\civ\ Narrow Line Emission}
\label{nc}

All tabulated parameter means and correlation coefficients
discussed above depend upon proper processing of the \civ\
spectra. Confusion exists about the reality and strength of a
narrow line \civ\ component (\civnc) presumably arising from the
same narrow-line region (NLR) as e.g. \oiiiopt\ and narrow \hb.
There is now no doubt that \civnc\ emission is common in AGN
\citep[see also][]{sulenticmarziani99}. High and low redshift type
2 AGN with obvious \civnc\ emission have recently been found in
significant numbers: \citet{bargeretal02};
\citet{jarvisetal05}(*); \citet{normanetal02}(*);
\citet{sternetal02}(*); \citet{szokolyetal04};
\citet{mainierietal05}(*); \citet{severgninietal06}. According to
\citet{meiksin06} only four confirmed high redshift ($z > $1.6)
type 2 AGN are known (refs. marked with * above). All four show
prominent \civnc\ \citep[see also][]{dawsonetal01}.


In contrast to \hb\ a clear, unique NLR/BLR inflection is less
often seen in the \civ\ profiles making NLR correction less
certain. This is not surprising when one considers that the
intrinsic velocity resolution at \civ\ is 3 times lower than at
\hb. NLR \civ\ can also be broader than other narrow lines
because: a) it is a doublet with $\Delta v  \approx $ 500 \kms\
and b) it can arise in denser than average parts of the NLR
\citep[as for \o4363; e.g. ][]{marzianietal96,sulenticmarziani99}.
We argue that cautious subtraction of a suitable narrow component
is essential for exploiting the information content in the \civ\
line \citep[see ][]{bachevetal04}. We subtracted a significant (W
$>$ 1 \AA) NLR component from 76/130 sources in this sample.
Figure \ref{fig:civbc} shows the individual \civ\ profiles with
narrow components indicated in order to assist visual assessment
of the component on a source-by-source basis.

\citet{baskinlaor05} recently pointed out that our earlier \civbc\
and \civnc\ measurements \citep{marzianietal96} were
``non-unique". Every experimental measure is non-unique and the
lack of uniqueness is customarily indicated by error bars. Rather
than uniqueness, the question we are addressing is whether or not
there is a significant narrow component in the \civ\ line. The
second question, assuming that such a component is present,
involves how accurately we can measure it. The third question,
assuming we can accurately measure it, is whether correction for
\civnc\ matters. The goals of this section are to provide a recipe
for consistent \civnc\ correction and to show that very different
results emerge from corrected \civ\ measures.

The strong and relatively narrow core (FWHM $\la$ 2000 \kms)
observed in many \civ\ profiles was previously noted and an {\it
ad hoc} intermediate line region (ILR) was introduced in order to
account for it \citep{brothertonetal94,brothertonfrancis99}. The
ILR was defined as having some properties typical of the canonical
BLR necessitating the postulation of an additional  VBLR component
in order to explain the broad wings often seen in \civ\ spectra
(e.g.  Figure \ref{fig:civbc}). Unfortunately the ILR approach is
not fully consistent because narrow \civ\ cores are significantly
narrower than corresponding \hbbc\ profiles
\citep{sulenticmarziani99} which are a canonical BLR feature. They
are sometimes as narrow as the  \oiii\ lines. Intermediate
ionization lines of \ciii\ and \siiii\ measured in average spectra
\citep{bachevetal04} show widths that are more consistent with
\hbbc\ and much broader than the narrow cores of \civ\ that we
ascribe to the NLR. In addition density-sensitive ratios measured
near the \ciii\ blend are consistent with BLR density (\ne\
 $ \sim 10^{10} $ \cm3: see \citet{brothertonetal94,bachevetal04}). This
reinforces our interpretation that the hypothesized ILR+VBLR
components as simply the more canonical NLR+BLR. The larger width
of \civnc\ compared to \hbnc\ or \oiii\ can be easily explained
within the framework of a density/ionization gradient within the
NLR, as further described below.

Almost all other studies of \civ\ line properties
\citep{willsetal95,corbor96,willsetal99,vestergaard02,warneretal04}
do not subtract \civnc\ emission. \citet{baskinlaor05} assume that
the width and strength of \civnc\ and \oiii\ are correlated.  In
most cases this implies that the ratio  \civ/\oiii\ is  $\simlt
1$.  The physical relationship between forbidden \oiiiopt\ and
permitted \civnc\ is however unclear leaving little basis for
assuming a fixed relation. The \oiiiopt\ lines often show a strong
blue wing that might be described as a semi-broad component.
So-called blue outlier sources  show this component and it is
expected to be a {\em strong} \civ\ emitter \citep{zamanovetal02}.
Our analysis suggests that \civnc\ is likely absorbed by dust or
is intrinsically weak in $\approx$ 50\% of sources. Among the
remainder about 1/3 of the sources show \civnc\ significantly
broader than narrow Balmer and \oiiiopt\ emission.  It is probably
emitted by a reddening-free high-density (or high-ionization)
innermost region of the NLR. Whatever its origin and relationship
to other narrow lines it is present in the spectra of many sources
and will affect our efforts to parameterize \civbc.

The motivation for relatively high density emission in the NLR
stems from the clear evidence of relatively large \civnc/\oiii\
intensity ratios in several sources: NGC 5548,
 NGC 7674 and I Zw 92 \citep{kraemeretal94,kraemeretal98}, with  \civnc/\oiii\
$\approx$ 2. Also, even if \citet{baskinlaor05} subtracted little
\civnc, the average non-zero subtraction for the 16 sources in
common with the present study implies \civnc/\oiii\ $\approx$ 0.3.
This value already indicates bulk emission from $\log$ \ne $\simgt
5.5$, much above the ``standard" NLR density \ne $\sim 10^4 $
\cm3. The \civ/\oiii\ intensity ratio increases with density
around the \oiiiopt\ critical density because of the drastic
collisional quenching that suppresses \oiiiopt\ but not \civ. The
observed FWHM differences  between \oiiiopt, \hbnc, and \civnc\
are recovered under standard assumptions if a density gradient is
assumed for the NLR, with $3 \simlt \log$ \ne $\simlt 7 \div 8  $
\citep{sulenticmarziani99}.


We suggest the following \civnc\ subtraction procedure as the most
reliable  way to obtain reasonable and reproducible \civnc\
measures.

\begin{description}

\item[Step 1: Inflection]  Sources showing a \civ\ NLR/BLR
inflection can be treated the same as \hb\ as long as the
width/shift/intensity constraints given below are not violated.
See the profiles in Figure \ref{fig:civbc} and Appendix A
discussion of PG 0026+126 which shows a strong profile inflection.
There was no simultaneous fitting. The underlying \civbc\ was fit
with an high-order spline function. The overlying narrow component
was set by bordering the fitting range at inflection points which
defined a core that met the FWHM and flux ratio criteria described
below. The FWHM was measured using a Gaussian fit, or by measuring
the half-maximum wavelengths if the profile was absorbed or very
different from Gaussian.

\item[Step 1a: No Inflection or Multiple Inflections] Most sources
do not show an inflection or sometimes show multiple inflections
between reasonable limits of width and strength. This motivates us
to set a conservative limit on FWHM \civnc. Simple models suggest
that lines like \civ\  can be significantly broader than \oiii\
\citep{sulenticmarziani99}. \civ\ lines with FWHM $\la$ 1500 \kms\
are now observed in higher redshift type-2 AGN (see above
references).  We therefore suggest subtracting a \civnc\ component
with FWHM $\leq$ 1500 \kms\ again subject to shift and intensity
constraints that follow (see Fig. \ref{fig:civnc}).  In just two
cases (3C 110 and 3C 273) our data suggested a somewhat broader
component but inclusion/exclusion of these few sources as
processed, or reduction of the NLR component to this limit will
not affect the main conclusions of this study. The choice was
usually to maximize the \civnc\ FWHM within the flux ratio
condition as described below. Any narrow feature with FWHM $\la$
900 \kms\ has no physical meaning. The feature we identify as
\civnc\ shows FWHM $\approx$ 1200 $\pm$ 300 \kms in 95\%\ of
sources with significant narrow emission (Fig. \ref{fig:civnc}).

\item[Step 2: Nebular Physics and Observations] There is no strong
upper limit for the expected \civ/\oiii\ intensity ratio in the
absence of internal dust extinction. Both high ionization and high
density can produce an arbitrarily large ratio
\citep{continiviegas01,kraemeretal98,baldwinetal95}. We adopt
\civ/\oiii\ $\approx$ 10, derived for the high-ionization region
of NGC 5548 \citep{kraemeretal98}, as a {\em strict} upper limit.
Using again observational results  as a guideline, we consider
Seyfert 1 sources in our sample that show prominent, unambiguous
\civnc\ (NGC sources, PKS 0518-45, and 3C 390.3).  We find a large
dispersion in the reddening-corrected \civnc/\oiii\ ratio with a
mean value $\approx$ 2 and a maximum $\approx$ 5 (NGC 3783).
Therefore we can safely regard an \civnc/\oiii\ intensity ratio
$\approx$ 5 as an observationally defined boundary. If this
condition is appropriate the ($A_{\mathrm B}$\ corrected)
distribution  of \civ/\oiii\ intensity ratios (shown in Fig.
\ref{fig:civnc}) does not pose any special challenge, including
the few sources for which 5 $\la$ \civnc/\oiii\ $\la 10$\ (with an
uncertainty of $\pm$ 50\%\ these sources are not significantly
above our adopted limit of 5).

\item[Step 3: NLR shift] In most sources the  \oiii\ and/or \hbnc\
profile centroid is used to define the rest frame of a source.
Limited available \hi\ and CO measures of host galaxy emission
support this definition except for a few extreme Pop. A (some but
not all formally NLSy1s) blue outlier sources. We use the peak of
\hb\ to define the source rest frame of blue outliers. The \civnc\
profile centroid (Table \ref{tab:civ}) agrees with the optically
defined rest frame in most sources. Ninety percent of our sources
show a \civnc\ centroid within  $\pm 400 $\kms.  This is
reasonable considering that the average FWHM(\civnc) =1120 \kms\
and that \civnc\ is strongly sensitive to S/N.  Shifts of several
hundred \kms\ are occasionally observed and may be due to: a) an
intrinsic \civnc\ blueshift, b) narrow-line absorption that
creates a spurious shift to the red (and, indeed, inspection of
Fig. \ref{fig:civbc} reveals that this is the case for most
sources where \civnc\ appears to be significantly redshifted) and
c) poor rest frame determination. However only 5 sources out of 29
with $|$\dvr$|$(\civnc) $\ga$ 300 \kms\ show a $\Delta z \approx
\pm $ 0.001.

\end{description}

Figure \ref{fig:civnc} summarizes our \civnc\ measures: line
luminosity distribution of \civnc\ (upper right); distribution of
\civnc/\oiii\ luminosity ratios (lower left); distribution of
\civnc\ FWHM measures (upper right);  distribution of \civnc\
measures\ in the line luminosity--FWHM plane (lower right).
Application of above procedures resulted in a subtracted NLR
component usually less than W(\civnc) $\approx$ 10 \AA\ but with a
few extreme cases usually low luminosity Seyfert 1's. RL sources
show the largest fraction of detectable \civnc\ components (0.71)
compared to 0.48 for RQ AGN. Our Pop. B sources show a slightly
larger fraction of \civnc\ components (0.63) than Pop. A (0.51).
Some sources do not allow an unambiguous \civnc\ subtraction with
a significant range of acceptable solutions. This ambiguity and
its effect on \civbc\ are usually within the adopted errors (even
if the effect on \civnc\ is much larger), that have been estimated
changing the fractional intensity levels by $\pm 5$\%. As
described earlier, the random scatter in Galactic line radial
velocity after realignment is just $\approx 40$ \kms.  Therefore
it is possible that several \civnc\ shifts are significant because
they show values larger than the expected calibration and
measurement uncertainties. Examining spectra in Fig.
\ref{fig:civbc} one will occasionally see a \civ\ profile with a
peak as narrow as some subtracted \civnc\ (e.g. J13253$-$3824 and
J15591+3501). In these cases subtraction of the sufficiently
narrow peak would violate other selection rules (e.g. in the above
two cases \civnc/\oiii $\gg$ 10). Note that we also verified {\it
a posteriori} that the \civnc\ FWHM was less than FWHM(\hbbc).


\subsection{The Narrow Cores of \civ\ Do Not Reverberate}

An ideal check on our NLR results would involve reverberation
mapping where any NLR component would be expected to remain
stable. One IUE based study \citep{turlercourvoisier98} reported
PCA analysis on 18 AGN with 15 or more independent spectra. Ten of
the sources are included in our sample. The principal component in
their study was interpreted to involve the parts of the \civ\ line
profile that varied with zero lag time. The approach of T\"urler
\& Courvoisier  was to isolate the principal component dominated
by continuum and broad line variability. This was then subtracted
from the mean spectrum to isolate the remaining information
content (rest spectrum). Two things are seen in the rest spectrum:
a narrow unshifted peak, and more complex and extended wings. The
nature of the wings will depend upon the complexity and timescale
of variations as well as the number and temporal spacing of source
spectra. Component 1 can be reasonably argued to be the NLR
component of the line -- the correlated intensity component that
dominated our 2D analogy above.

In the case of 3C 273 only the continuum was present in the
principal component. We identified and subtracted an NLR component
in all ten overlap cases. A narrow component of similar strength
and width is seen in the second principal component spectra for
nine of these cases (except 3C 273). The least ambiguous case
involves 3C 390.3 where there is a clear inflection between NLR
and BLR. In that case agreement is perfect. Other sources like GQ
Com, NGC 3783 and NGC 5548 also show strong agreement. The range
of FWHM for the second principal component \civ\ profiles 1-5000
\kms\ suggesting that the NLR is often blended with additional
broad line emission. However, the overall agreement between the
central cores and our own estimates of NLR \civ\ emission gives us
confidence that we have developed a reasonable approach to
correcting the \civ\ line profiles. The alternative is to ignore
the problem which we argue will lead to spurious results.


\subsection{Comparison with Previous Work}

Other recent studies of the \civ\ profile, using the same HST
archival spectra, subtracted little \citep{baskinlaor05} or no
\citep{willsetal93,corbor96,vestergaard02,kuraszkiewicz02,kuraszkiewicz04,warneretal04}
NLR component. Figure \ref{fig:bls} compares our \civbc\ FWHM and
centroid shift ($c(\case{1}{2})$) measures with equivalent values
for sources in common with some of these studies.  The LL panel of
Figure \ref{fig:civnc} shows that \citet{baskinlaor05} subtracted
a ($\sim$2-4 times) smaller and more constant NLR component.
Direct comparison with \citet{kuraszkiewicz02,kuraszkiewicz04} is
not possible because they model the \civ\ profile with multiple
Gaussian components that do not correspond to our NLR and BLR
interpretation. FWHM measures are strongly affected by
under-subtraction of \civnc.  The UL panel of Figure \ref{fig:bls}
compares our FWHM \civbc\ measures with those of
\citet{baskinlaor05} and \citet{warneretal04}.  Symbols for
comparisons with \citet{baskinlaor05} (and \citet{corbor96})
retain the  Pop. A-B and RQ/RL distinctions used in earlier
figures. Our measures are systematically larger with  the
amplitude of $\Delta$ FWHM increasing systematically with FWHM
\civbc. The LL panel compares our FWHM measures with
\citet{corbor96} and shows similar disagreement. Correlations such
as FWHM \hb\ vs. FWHM \civ\
\citep{corbin91,baskinlaor05,warneretal04} found using uncorrected
\civ\ measures will likely be spurious except possibly for the
Pop. B sources. The most striking evidence for correlation is
found in Figure 7 (right) of \citet{corbor96} involving NC
corrected \hbbc\ and uncorrected \civbc\ measures. One sees two
groups of sources (pop. A and B) each showing a positive trend but
with different slopes for the two trends. The trends are displaced
by $\Delta$ FWHM(\civbc)=3000\kms at about FWHM(\hbbc) = 4000\kms.
The ``Pop. B'' trend can be described as displaced towards smaller
values of FWHM(\civbc). Since narrow line emission is
systematically stronger in Pop. B (especially RL) sources we might
expect those FWHM(\civbc) measures to be more strongly affected by
NC subtraction. Is the displacement entirely due to the lack of NC
corrected FWHM(\civbc) measures? Much of the displacement
disappears in our equivalent FWHM-FWHM plot but the correlation
seen for Pop. B sources ($r_\mathrm{S} \approx$ 0.5) is stronger
than for Pop. A ($r_\mathrm{S}  \sim$0.3: not significant) and its
extrapolation into the pop. A domain predicts much smaller (3000
\kms) values for FWHM(\civbc) than are observed.

The right panels of Fig.\ref{fig:bls} compares our
$c(\case{3}{4})$ measures with those from \citet{baskinlaor05}
(upper) and \citet{corbor96} (lower).  There is a systematic
displacement of uncorrected shift measures towards smaller or even
redshifted values in both comparisons. This will tend to diminish
the Pop. A-B (or RQ vs. RL) differences that are highlighted in
this paper. The systematic \civ\ blueshift for Pop. A sources
becomes much less obvious using NC uncorrected \civ\ measures and
especially using shift measures taken closer to the profile peak
(e.g., $c(0.9$)). Fig. \ref{fig:bls} shows systematic differences
between corrected and uncorrected measures that will erase or
obscure important  \civ\ results like the ones discussed in this
paper.

\section{\mbh\ Calculations Using \civ\ Width }
\label{mbh}

\civ\ has become the line of choice for black hole mass estimation
in high- $z$\ quasars. It is a dangerous choice for at least two
reasons: 1) it shows a systematic blueshift in many sources, and
2) FWHM \civbc\ does not correlate strongly or monotonically with
FWHM \hbbc\ -- the line of  choice for low-redshift \mbh\
estimation. Reason 1 does not necessarily rule out a virialized
distribution of emitting clouds but it certainly motivates caution
when using the line to infer black hole mass. Blueshifted \civ\
profiles are thought to arise in a high ionization wind resulting
in a velocity flow that is not negligible relative to any
rotational component \citep{murrayetal95,progakallman04}. We think
use of \civ\ warrants even more caution because we see different
line properties for Pop. A and B (or alternatively  RQ and RL)
sources.  This raises the possibility that the geometry/kinematics
of the \civ\ emitting region may be fundamentally different in
Pop. A and B sources. The population distinction is at least
useful and possibly fundamental because it maximizes source
differences. FWHM \hbbc\ and \civbc\ are most similar (Table
\ref{tab:ave}) for RQ sources that show  mean FWHM(\civbc) only
$\approx$600 \kms\ larger than FWHM(\hbbc). The RQ source
distinction will therefore yield reasonable agreement between the
two \mbh\ estimators. The same is true for sources under the RL
distinction where FWHM(\civbc) is  $\approx$ 900 \kms\ broader.
Both differences are approximately 15-16\% of the mean RQ and RL
profile widths respectively.

The two lines show larger difference when sources are divided
using the Pop. A-B distinction where $\Delta$ FWHM (\hbbc)  --
FWHM(\civbc) $\approx$ --1900 \kms\ and $\approx +1400$ \kms\ for
pop. A and B respectively. These discrepancies amount to $\sim $56
\%\ and $\approx $ 21 \%\ of FWHM(\civbc) + FWHM(\hbbc)/2 for pop.
A and B, respectively. This is larger than the measurement
uncertainties for FWHM measures of both lines and further supports
the utility of the pop. A-B concept. The two estimators will yield
\mbh\ estimates that are more discrepant. Adopting the Pop. A-B
distinction as more useful than the RQ-RL one then finds the
largest Pop. A-B differences using FWHM(\hbbc) where $\Delta$FWHM
(A-B) $\approx$ -4600 \kms\ compared to --1400 \kms\ using
FWHM(\civbc).  The corresponding differences for the RQ-RL
distinction are $\Delta$ FWHM (RQ--RL) $\approx$ --2800\kms\
(\hbbc) and --1200 \kms\ (\civbc).

As already pointed out \citep[e. g., ][]{marzianietal96} the
intrinsic dispersion of FWHM \civbc\ is less than for FWHM\hbbc\
making it less sensitive to differences between source
populations. Since FWHM \civbc\ measures are less accurate than
FWHM \hbbc\ values derived BH masses using the former will blur
out any trends obtained with \hbbc\ measures. FWHM \civbc\
-derived masses will yield much larger \mbh\ estimates for Pop. A
and smaller values for Pop. B. If one prefers to avoid the Pop.
A-B distinction then one will find smaller \civbc\ -- \hbbc\
differences using the RQ-RL distinction perhaps encouraging the
incorrect assumption that a simple correlation exists between the
two sets of \mbh\ measures. The smaller difference between mean
FWHM values has also caused some to conclude that RQ and RL
sources have similar \mbh\ distributions and mean values. Even if
FWHM(\civbc) could be measured with equal accuracy, and confidence
about viriality, as FWHM(\hbbc) it would be a less useful \mbh\
estimator because it shows less dispersion. The main source of
disagreements about \mbh\ similarities and differences among AGN
samples involves Pop. B RQ sources.  Combining them with narrower
lined RQ sources will raise the mean value of \mbh\ for that
population with only small affect on the RL results.  It will tend
to equalize the means.

Estimates of \mbh\ were obtained from the UV flux density and FWHM
\civbc\ reported in Table \ref{tab:civ}, as well as for the
corresponding data from \citet{baskinlaor05} (their Table 1),
assuming Hubble constant $H_0 = 70 $ \kms\ Mpc$^{-1}$ and relative
energy density $\Omega_\Lambda = 0.7$\ and $\Omega_{\mathrm M} =
0.3$. Values of \mbh\ were derived following the latest
normalization of \citet{vestergaardpeterson06}, which use the same
cosmological parameters.  The upper panel of Figure \ref{fig:mbh}
compares \civ\ based \mbh\ estimates of \citet{baskinlaor05}
(based on slightly \civnc- corrected \civ\ measures) with the
NC-corrected estimates derived from this paper. We see that
\citet{baskinlaor05} measures are systematically low and that the
difference from our results increase with \mbh. This comparison
involves only sources in common between the two studies and
involves only a 2dex range in \mbh. The differences between our
measures and completely uncorrected \civ\ profiles will be larger.
We note that both Pop. A and B sources show this disagreement. The
middle panel of Fig. \ref{fig:mbh} compares \mbh\ measures based
upon FWHM \civbc\ and \hbbc. We show the ratio of \civbc/\hbbc\
--derived \mbh\ measures as a function of \hbbc--derived \mbh. The
\hbbc\ and continuum flux density measures come from
\citet{marzianietal03a}. The latest normalization of
\citet{vestergaardpeterson06} was applied to these data, too.

We suggest a corrected FWHM(\hbbc) measure (reduced by a fraction
dependent on FWHM(\hbbc)) as likely to be the most reliable virial
estimator for reasons described in \citet{sulenticetal06a}. The
middle panel of Fig. \ref{fig:mbh} suggests that (NC corrected)
\civ\ based \mbh\ estimates for Pop. B sources are more consistent
with ones computed from the corrected \hbbc\ width. However the
scatter is large  and our \civ\ regression line is 0.2dex higher
than for \mbh\ derived from \hbbc. The most serious disagreement
involves Pop. A sources ($\approx$ 60\%\ of RQ sources) which show
a trend where the \mbh\ ratio increases with decreasing \mbh. This
does not allow one to easily correct \civ-computed \mbh\ to \hbbc\
values unless information on the optical spectrum (rest frame and
\hbbc\ line width) is available. We made several attempts to
deduce a correction for \civ-derived \mbh\ values from properties
intrinsic to the \civ\ profile shape (i.e., width, asymmetry and
kurtosis) but were unable to find an effective relationship.
 Perhaps the most effective relationship we found
involves the one shown in the lower panel of Fig. \ref{fig:mbh}
which shows that the ratio of \mbh\ derived from \civ\ and \hbbc\
is loosely correlated with W(\civbc) (for W(\civnc) $\la $ 100
\AA). \civbc\ and \hbbc\ estimates  of \mbh\ show better agreement
for larger values of  W(\civbc). Caution is advised because
equivalent width measures may be affected by continuum reddening.
We also suffer from a relatively small sample of sources with
W(\civbc) $\ga$ 100 \AA.  Our fears about \civ--derived estimates
for \mbh\ have motivated us to pursue \hb\ to the highest possible
redshift and we have recently presented \hbbc\ derived \mbh\
estimates out to $z \approx $ 2.5
\citep{sulenticetal04,sulenticetal06a}.

If NLR \civ\ follows \oiii\ then we expect the strongest and most
frequent \civ\ NLR to affect Pop. B sources. Uncorrected \civ\
profiles in Pop. B sources will then be measured with
systematically narrow FWHM \civ. Using this measure for \mbh\
estimation will result in systematic under estimation. This
explains why many of previous studies found little or no
difference in \mbh\ estimates for RQ and RL sources in direct
contradiction with derivations based on FWHM \hbbc.  The overall
tendency will be to push both ends towards the center thus
reducing the dispersion of \mbh\ derived from uncorrected \mbh\
\citep[see also discussion in][]{baskinlaor05}. \mbh\
underestimates for many sources, especially RL which tend to be
overluminous in an optically selected sample (e.g. BG92), will
yield spuriously high \lledd\ values  \citep[e. g.
][]{warneretal04} for Pop. B. The insidious effect of uncorrected
\civ\ measures is that it will tend to mix sources of very
different empirical and physical properties. The correlation
between FWHM \civ\ and \lledd\ \cite[][ Fig. 5]{warneretal04} is
almost certainly driven by biases resulting from use of
uncorrected \civ\ measures.

\section{Conclusions}

AGN are widely compared and contrasted in two ways: (1) RQ vs. RL
and (2) NLSy1 vs. broad-line Seyferts/quasars. We suggest an
alternate approach that unites both of these distinctions and that
is supported by differences in \civ\  line measures. We find that
sources above and below FWHM \hbbc\ $\approx$ 4000 \kms\ show the
most significant spectroscopic (and broadband) differences. RL
sources lie largely above this limit while NLSy1 lie below it. We
find that all or most sources below 4000 \kms\ show properties
similar to NLSy1s. Figure \ref{fig:e1} (upper left panel) in this
paper particularly reenforces this similarity by showing that
almost all sources with FWHM \hbbc $\la$ 4000 \kms\ show a
systematic \civ\ blueshift. Our population A-B concept simply
reflects a unification where Pop.  A sources show NLSy1-like
properties and Pop. B sources show RL-like properties.

This paper addresses two thorny problems involving \civ\ measures
and their interpretation: (1) when and how to correct \civ\ for
NLR contamination and (2) whether \civ\ measures support previous
claims, based on optical spectra (and radio loudness), for two
Populations (A+B) of broad line AGN. The second result actually
clarifies the answer to the first problem. Evidence is now
ubiquitous at both high and low redshift for significant \civ\ NLR
emission in many sources. If we used \oiii\ as a line template
then we would find fewer \civnc\ components and those found would
be narrower and weaker (i.e. lower EW). In several cases we find
such an \oiii-like component but in many sources our inferred
\civnc\ component is broader and hence stronger. We argue that
empirical evidence (e.g., inflections in some sources and broader
\civnc\ in Type 2 AGN) as well as simple models support our
hypothesis that \civnc\ is often not ``\oiii-like."

We argue that correlations found without NLR correction are very
often spurious while real correlations (Figs. \ref{fig:e1},
\ref{fig:civ}) require NLR correction to be seen clearly. We
propose a simple recipe for \civ\ NLR correction. \civ\ proves to
be a valuable 4DE1 space discriminator that provides evidence in
support of our two population hypothesis in the sense that the
\civ\ blueshift is ubiquitous only in previously defined
population A sources. These results have strong implications for
any attempt to use \civ\ measures for black hole mass (and \lledd)
estimation. We suggest that any use of \civ\ line measures can be
facilitated by interpreting them within the 4DE1 + population A-B
context.


We thank the referee for thorough readings of the manuscript. DD
acknowledges support from grant IN100703 PAPIIT UNAM. This
research has made use of the NASA/IPAC Extragalactic Database
(NED) which is operated by the Jet Propulsion Laboratory,
California Institute of Technology, under contract with the NASA.
Funding for the SDSS and SDSS-II has been provided by the Alfred
P. Sloan Foundation, the Participating Institutions, the National
Science Foundation, the U.S. Department of Energy, the National
Aeronautics and Space Administration, the Japanese Monbukagakusho,
the Max Planck Society, and the Higher Education Funding Council
for England. The SDSS Web Site is http://www.sdss.org/.  The SDSS
is managed by the Astrophysical Research Consortium for the
Participating Institutions. The Participating Institutions are the
American Museum of Natural History, Astrophysical Institute
Potsdam, University of Basel, University of Cambridge, Case
Western Reserve University, University of Chicago, Drexel
University, Fermilab, the Institute for Advanced Study, the Japan
Participation Group, Johns Hopkins University, the Joint Institute
for Nuclear Astrophysics, the Kavli Institute for Particle
Astrophysics and Cosmology, the Korean Scientist Group, the
Chinese Academy of Sciences (LAMOST), Los Alamos National
Laboratory, the Max-Planck-Institute for Astronomy (MPIA), the
Max-Planck-Institute for Astrophysics (MPA), New Mexico State
University, Ohio State University, University of Pittsburgh,
University of Portsmouth, Princeton University, the United States
Naval Observatory, and the University of Washington.

\clearpage

\appendix
\section{Notes on Individual Objects}
\label{appendix}

Most sources follow the 4DE1 trends described here and in previous
papers. However a few sources appear to be genuinely pathological.
We mention a couple of such sources that appear as outliers in
4DE1 space and that are particularly relevant to the discussion
involving \civ\ measures.

\begin{description}

\item[3C 57] shows $c(\case{1}{2})$\ \civ\ and \rfe\ parameters
typical of a Pop. A (even extreme Pop. A, NLSy1s) source
($c(\case{1}{2})$ = --1605 \kms; \rfe $\approx$ 1.25). W(\civ) and
profile shape are also typical of Pop. A (even similar to the ones
of I Zw 1. At the same time it is RL, shows no soft X-ray excess
and FWHM(\hbbc) $\approx$ 4700 \kms\ all consistent with Pop. B.

\item[PG 0026+126] This quasar is moderate RQ Pop.  A following
the current 4DE1 empiricism because FWHM(\hbbc) $\approx$ 2400
\kms\ and \rfe $\approx$ 0.28. There are two possible
interpretations of the \civ\ profile: (1) (FWHM \civ $\approx$
1860 \kms\ and $c(\case{1}{2})$ = +140 \kms) if the strong narrow
peak is {\em not} subtracted or 2)(FWHM \civ $\approx$ 7000 \kms\
and $c (\case{1}{2}) \sim$-1000 \kms) if the narrow peak is
subtracted as a NLR component. This last approach seems especially
appropriate since FWHM of the  \civ\ narrow core only slightly
exceeds (10--20 \%) FWHM(\oiii). The source is a FWHM \civ\
``outlier" whichever \civ\ measure is adopted --either unusually
narrow or unusually broad (see Figure \ref{fig:civ}). \rfe\ and
\gs\ measures are intermediate for the source and therefore
unconstraining. Note that an erroneous rest frame is often assumed
for this source \citep{geldermanwittle94}. The most accurate
redshift for the source corresponds to the centroid of the narrow
component which is consistent with the NLR interpretation. This
also yields a modest blueshift for the broader component which is
also unconstraining. The strong profile inflection and  small FWHM
for the unshifted narrow component lead us to subtract it as NLR
emission.  RQ sources like NGC 4253 and 4395 show similar
NLR-strong profiles.

\item[PKS 1252+119] Is the highest-$z$ \ quasar in our sample.
\hb\ is consequently located at the edge of an excellent SDSS
spectrum, making  measures of FWHM(\hbbc) uncertain. The reported \rfe\
is the only upper limit in our sample (marked with an arrow in
Fig. \ref{fig:e1}). This source maybe  located in an area of
the 4DE1 optical plane where other core-dominated RL sources are
found \citep[Fig. 1 of ][]{sulenticetal03} but confirmatory
optical data are needed.
\end{description}


\begin{deluxetable}{llrcccccccc}\tabletypesize{\scriptsize}
\tablecomments{Table 1 is published in its entirety in the
electronic edition of the Astrophysical Journal. A portion is
shown here for guidance regarding its form and content.}
\tablenum{1} \tablewidth{0pt} \tablecaption{Source Identification
and 4DE1 Parameters} \tablehead{\colhead{IAU Code}&
\colhead{Common Name\tablenotemark{a}}&
\colhead{$z$\tablenotemark{b}} &  \colhead{$z$\ Ref.} &
\colhead{PG/BO} &\colhead{$A_{\mathrm B}$\tablenotemark{c}} &
\colhead{FWHM(\hbbc)} & \colhead{\rfe} & \colhead{$\log$(\rk)} & \colhead{$\Gamma_{\rm soft}$} \\
& & & & & [mag]  &   [\kms] &  & & \\
\colhead{(1)} & \colhead{(2)} & \colhead{(3)} & \colhead{(4)} &
\colhead{(5)} & \colhead{(6)} & \colhead{(7)} & \colhead{(8)} &
\colhead{(9)}& \colhead{(10)} } \startdata
J00057+0203   & LBQS 0003+0146          & 0.234     &  C91     &      &0.10 &3315   & 0.26 &   -1.00   &3.08    \\%
J00059+1609   & PG 0003+158             & 0.4504    &  M03     &$\ast$&0.22 &5519   & 0.14 &   2.53    &2.34     \\%
J00063+2012   & MRK 0335                & 0.0252    &  M03     &$\ast$&0.15 &1950   & 0.28 &   -0.54   &2.90     \\%
J00204+0226   & LBQS 0017+0209          & 0.401     &  F89     &      &0.10 &2535   & 1.07 &   -1.00   &\ldots       \\%
J00292+1316   & PG 0026+129             & 0.1451    &  M03     &$\ast$&0.31 &2405   & 0.28 &   -0.10   &2.07       \\%
J00392$-$5117 & [WPV85] 007             & 0.0290    &  G99     &      &0.05 &1203   & 0.68 &   \ldots  &9.00:     \\%
J00449+1026   & [HB89] 0042+101         & 0.5857    &  SPM     &      &0.29 &8978   & 0.40 &   0.00    &2.48        \\%
J00470+0319   & PG 0044+030             & 0.6231    &  M03     &      &0.09 &5759   & 0.19 &   1.83    &1.72  \\
\enddata
\tablenotetext{a}{In a format recognized by NED.}
\tablenotetext{b}{Accuracy of $z$ values can be in general assumed
to be $\pm 0.0001$\ at a 1--$\sigma$\ confidence level in case
four decimal digits are provided; $\pm 0.001$\ otherwise. }
\tablenotetext{c}{From \citet{schlegeletal98}} \tablerefs{M03:
\citep{marzianietal03a}; G99: \citet{grupeetal99}   SDSS: Spectra
retrieved from www.sdss.org. $z$\ values were measured as
described in the text of the paper, and may differ from those
reported in NED. L67: \citet{lynds67}. SPM: Unpublished spectra
obtained with the 2.2m telescope at San Pedro Martir. B96:
\citet{brotherton96}. ESO: Unpublished ESO spectra. KPNO:
unpublished KPNO spectra.  T93: \citet{tadhunteretal93}.  JB91:
\citet{jacksonbrowne91} C97: \citet{corbin97}. M96:
\citet{marzianietal96}. LBQS:  W00: \citet{wisotzkietal00}  A91:
K96: \citet{keel96}.   EH04: \citet{eracleoushalpern04}. A91:
\citet{allenetal91}. HB89: \citet{hewittburbidge89}. HB80:
\citep{hewittburbidge80}. }
\end{deluxetable}

\clearpage

\textwidth=20cm\hoffset=-1cm
\begin{deluxetable}{llrrrrrrrrrrrrrrrrrrrccccc}
\rotate \hoffset=-2.5cm \voffset=3.25cm
\tabletypesize{\tiny} \tablecomments{Table 2 is published in its
entirety in the electronic edition of the Astrophysical Journal. A
portion is shown here for guidance regarding its form and
content.} \tablenum{2} \tablewidth{0cm} \tablecaption{\civ\
Emission Line Parameters} \tablehead{\colhead{IAU Code}&
\colhead{$f_\lambda$\tablenotemark{a}}& \colhead{$F_{\mathrm
NC}$\tablenotemark{b}} & \colhead{\dvr$_{\mathrm
NC}$\tablenotemark{d}} & \colhead{$F_{\mathrm
BC}$\tablenotemark{c}} &
\colhead{$c$($\case{1}{4}$)\tablenotemark{d}} &
\colhead{$\Delta^-$}\tablenotemark{d} &
\colhead{$\Delta^+$}\tablenotemark{d} &
\colhead{$c(\case{1}{2})$\tablenotemark{d}} &
\colhead{$\Delta^-$\tablenotemark{d}} &
\colhead{$\Delta^+$\tablenotemark{d}} & \colhead{$c$(3/4)} &
\colhead{$\Delta$\tablenotemark{d}}   &
\colhead{c(0.9)\tablenotemark{d}} &
\colhead{$\Delta$\tablenotemark{d}} &
\colhead{FWHM\tablenotemark{d}} &
\colhead{$\Delta$\tablenotemark{d}}  & \colhead{A.I.} &
\colhead{$\Delta^-$}& \colhead{$\Delta^+$}  & \colhead{$Kurt.$}  &
\colhead{$\Delta$}\\
\colhead{(1)} & \colhead{(2)} & \colhead{(3)} &
 \colhead{(4)} &
\colhead{(5)} & \colhead{(6)} & \colhead{(7)} & \colhead{(8)} &
 \colhead{(9)} &
\colhead{(10)} & \colhead{(11)} & \colhead{(12)} & \colhead{(13)}
& \colhead{(14)} & \colhead{(15)} & \colhead{(16)} &
\colhead{(17)} & \colhead{(18)} & \colhead{(19)} & \colhead{(20)}
& \colhead{(21)} & \colhead{(22)} }\startdata
J00057+0203 &   0.55  &   0.23    &   -726      &  4.0 &   -824    &   331 &   560 &   -455    &   121 &   120 &   -404    &   110 &   -365    &   78  &   4168    &   252 &   -0.12   &   0.10    &   0.16    0.16    &   0.35    &   0.06    \\%
J00059+1609 &   2.53  &   1.94    &   -66       &  22.5    &   866 &   591 &   522 &   35  &   200 &   228 &   -8  &   143 &   -1  &   92  &   5347    &   455 &   0.16    &   0.12    &   0.10    0.12    &   0.29    &   0.04    \\%
J00063+2012 &   7.97  &   7.33    &   314       &  50.7    &   -221    &   259 &   229 &   19  &   98  &   94  &   147 &   81  &   195 &   53  &   2927    &   195 &   -0.17   &   0.11    &   0.10    0.11    &   0.37    &   0.05    \\%
J00204+0226 &   0.44  &   0.00    &   \nodata   &  1.2 &   -2426   &   174 &   155 &   -2144   &   122 &   117 &   -2063   &   140 &   -2064   &   106 &   5378    &   244 &   -0.10   &   0.07    &   0.07    0.07    &   0.50    &   0.05    \\%
J00292+1316 &   2.82  &   1.64    &   95        &  8.5 &   1035    &   1077    &   244 &   -501    &   199 &   215 &   -201    &   261 &   -19 &   160 &   8595    &   429 &   0.14    &   0.17    &   0.06    0.17    &   0.36    &   0.06    \\%
J00392-5117 &   0.20  &   A       &   \nodata   &  2.0 &   338 &   262 &   265 &   -78 &   108 &   186 &   -54 &   81  &   -6  &   55  &   3056    &   371 &   0.11    &   0.09    &   0.09    0.09    &   0.28    &   0.03    \\%
J00449+1026 &   0.12  &   0.18    &   56        &  2.3 &   1433    &   846 &   705 &   -37 &   268 &   281 &   -124    &   203 &   -144    &   125 &   7385    &   563 &   0.20    &   0.10    &   0.09    0.10    &   0.26    &   0.04    \\%
J00470+0319 &   1.09  &   0.22    &   126       &  8.2 &   -418    &   322 &   490 &   -750    &   203 &   229 &   -568    &   143 &   -591    &   91  &   6274    &   458 &   0.03    &   0.07    &   0.10    0.10    &   0.33    &   0.04    \\
\enddata
  \tablenotetext{a}{Specific continuum flux at 1550 \AA\ in units of \ergss\ \AA$^{-1}$ cm$^{-2} \times 10^{14}$. }
\tablenotetext{b}{Flux of \civnc\ in units of \ergss\ cm$^{-2}
\times 10^{13}$. The letter ``A" indicate major absorptions
(typically of mini-BALs) affecting the profile of \civbc. ``a"
indicates narrow absorptions that ``eat away" \civnc\ but that do
not hamper \civbc\ measurements.}\tablenotetext{c}{Flux of \civbc\
in units of \ergss\ cm$^{-2} \times 10^{13}$.
}\tablenotetext{d}{In units of \kms.}
\label{tab:civ}
  \end{deluxetable}

\clearpage

\begin{deluxetable}{cccccccc}
\tabletypesize{\scriptsize}  \tablenum{3} \tablewidth{0pt}
\tablecaption{{\sc Average Values\tablenotemark{a}}} \tablehead{
\colhead{AGN pop} & \colhead{$N_{\rm sources}$}
&\colhead{EW(\civbc)}& \colhead{$c(\case{1}{2}$) \civbc\ } &
\colhead{FWHM(\civbc) } & \colhead{FWHM(\hbbc) } & \colhead{\rfe}
& \colhead{$\Gamma_{\rm Soft}$}
\\
\colhead{} & & \colhead{[\AA]} & \colhead{[\kms]} &
\colhead{[\kms]} & \colhead{[\kms]} & \colhead{}  & \colhead{} \\
\colhead{(1)} & \colhead{(2)} & \colhead{(3)} & \colhead{(4)} &
\colhead{(5)} & \colhead{(6)} & \colhead{(7)} & \colhead{(8)} }
\startdata
All QSO  & 130  &     93  (66) & $-$294  (837)  &5284 (1787) &5387 (3268)&0.36 (0.38) &2.41 (0.58)\\ %
RQ  only &  71  &     84  (69) & $-$582  (860)  &4733 (1482) &4100 (2681)&0.48 (0.43) &2.64 (0.63)\\ %
RL  only &  59  &    104  (60) & $+$52   (667)  &5946 (1905) &6936 (3259)&0.22 (0.25) &2.15 (0.38)\\ %
Pop. A   &  52  &     57  (34) & $-$677  (966)  &4451 (1269) &2604 (891) &0.60 (0.45) &2.67 (0.68)\\ %
Pop. B   &  78  &    117  (71) & $-$39   (627)  &5839 (1871) &7242 (2943)&0.20 (0.22) &2.24 (0.43)\\  %
\enddata
\label{tab:ave} \tablenotetext{a}{The number in parenthesis
indicates the sample standard deviation.}
\end{deluxetable}

\clearpage


\clearpage

\begin{deluxetable}{cccccccccccccccccc}
\rotate    \tabletypesize{\scriptsize} \tablenum{4}
\tablewidth{0pt} \tablecaption{{\sc Correlation Analysis for
\civbc\ Parameters\tablenotemark{a}}} \tablehead{ \colhead{\civbc\
} & \multicolumn{2}{c}{W}  && \multicolumn{2}{c}{FWHM} &&
\multicolumn{2}{c}{$c(\case{1}{2})$/W} &&
\multicolumn{2}{c}{FWHM(\hbbc)} && \multicolumn{2}{c}{\rfe} &&
\multicolumn{2}{c}{\gs} \\ \cline{2-3} \cline{5-6} \cline{8-9}
\cline{11-12} \cline{14-15} \cline{17-18}
 &
$r_{\mathrm S}$ & $P_{r_\mathrm{S}}$ && $r_{\mathrm S}$ &
$P_{r_\mathrm{S}}$ && $r_{\mathrm S}$ & $P_{r_\mathrm{S}}$ &&
$r_{\mathrm S}$ & $P_{r_\mathrm{S}}$ && $r_{\mathrm S}$ &
$P_{r_\mathrm{S}}$ && $r_{\mathrm S}$ & $P_{r_\mathrm{S}}$  \\
& \colhead{(1)} & \colhead{(2)} && \colhead{(3)} & \colhead{(4)}
&& \colhead{(5)} & \colhead{(6)} && \colhead{(7)} & \colhead{(6)}
&& \colhead{(9)} & \colhead{(10)} && \colhead{(11)} &
\colhead{(12)} } \startdata
\multicolumn{18}{c}{All Sources\tablenotemark{b}}\\
\hline
\\
$c(\case{1}{2}$)        &   0.411   &   5.951E-07   &&  0.008   &   4.6E-01 &&  0.970   &   8.7E-81 &&  0.331   &   6.0E-05 &&  -0.431  &   1.5E-07 &&  -0.327  &   1.5E-04 \\
W                       &   \nodata &   \nodata &&  0.141   &   5.5E-02 &&  0.485   &   2.5E-09 &&  0.469   &   9.1E-09 &&  -0.530  &   4.5E-11 &&  -0.284  &   9.1E-04 \\%
FWHM                    &   \nodata &   \nodata &&  \nodata &   \nodata &&  0.040   &   3.3E-01 &&  0.512   &   2.4E-10 &&  -0.175  &   2.3E-02 &&  -0.354  &   4.2E-05 \\%
 $c(\case{1}{2}$)/W &   \nodata &   \nodata &&  \nodata &   \nodata &&  \nodata &   \nodata &&  0.381   &   3.9E-06 &&  -0.471  &   7.8E-09 &&  -0.336  &   9.9E-05 \\%
\\ \hline
\multicolumn{18}{c}{RQ \tablenotemark{c}}\\
\enddata\label{tab:cc}
\tablenotetext{a}{The number $P$\ in parenthesis yields the
probability that the coupled variables are not correlated. A
significant correlation can be assumed if $P \la 0.01$.}
\tablenotetext{b}{130 sources of which 118  with available \gs.}
\tablenotetext{c}{71 sources of which 62  with available \gs.}
\tablenotetext{d}{59 sources of which 56  with available \gs.}
\tablenotetext{e}{52 sources of which 46  with available \gs.}
\tablenotetext{f}{78 sources, of which 72 with available \gs.}
\end{deluxetable}

\clearpage

\begin{deluxetable}{lccl}
\tabletypesize{\scriptsize} \tablenum{5} \tablewidth{0pt}
\tablecaption{{\sc Main Trends Along the 4DE1 Sequence}}
\tablehead{\colhead{Parameter} & \colhead{Population A}
&\colhead{Population B}& \colhead{References}} \startdata
FWHM(\hbbc)                   & 800 -- 4000 \kms\ & 4000 -- 10000 \kms\ & 1,2,3,4,5\\
$\bullet$log \mbh\            & 6.5 -- 8.5       & 8.0 -- 10.0           & 6,7,8,9\\
$\bullet$ \lledd\             & 0.1 -- 1.0        & 0.01 -- 0.5        & 6,7,8,9\\
W(\hbbc)                      & $\sim$ 80 \AA\   & $\sim$ 100 \AA\     & 1,2,3,4\\
\hbbc\ profile shape          & Lorentzian       & double Gaussian    & 2,3,5,10,11\\
$c(\case{1}{2}$) \hbbc\       & $\sim$ zero      & +500 \kms\         & 1,2,3\\
\rfe\                         & 0.7              & 0.3                & 1,2 \\
{Si}{\sc iii} / {C}{\sc iii}] & 0.4              & 0.2                & 12,13\\
\enddata
\tablerefs{1: \citet{marzianietal96} 2: \citet{sulenticetal00a} 3:
\citet{sulenticetal00b} 4: \citet{marzianietal03a} 5:
\citet{veroncettyetal01}  6: \citet{kaspietal00} 7:
\citet{marzianietal03b} 8: \citet{petersonetal04} 9:
\citet{sulenticetal06a}; 10: \citet{sulenticetal02}  11:
\citet{sulenticetal00c} 12: \citet{marzianietal01} 13:
\citet{willsetal99} 14: \citet{yuanwills03}   15:
\citet{baskinlaor05} 16: \citet{wangetal96} 17:
\citet{turneretal99} 18: \citet{grupeetal01} 19:
\citet{giveonetal99} 20: \citet{sulenticetal03} 21: Sulentic et
al. (2007), in preparation 22: \citet{reichardetal03}
 22: \citet{sulenticetal06b}    } \label{tab:summ}
\end{deluxetable}

\clearpage

\begin{figure}
\plotone{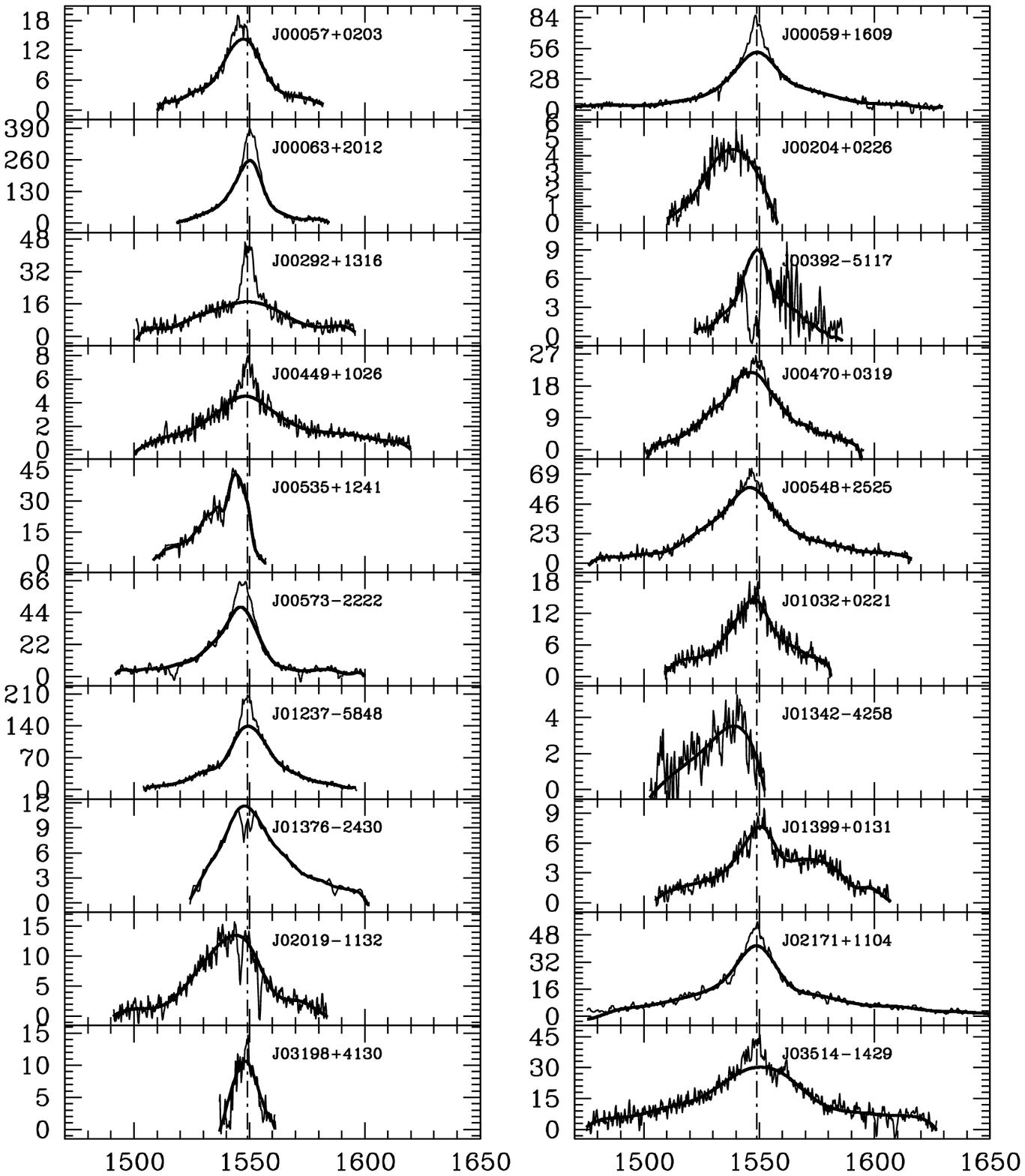} \caption{\civ\ profiles for the 130 AGN used in
the present study. They are presented in right ascension order.
Abscissa is rest-frame wavelength (\AA); ordinate is rest-frame
specific flux (10$^{-15}$ \ergss cm$^{-2}$ \AA$^{-1}$). Thick
curves are a high-order spline fits to \civbc\ and anything above
it is considered to be narrow line emission. The major thick
spacing (50 \AA) corresponds to a radial velocity range of \dvr
$\approx$ 9700 \kms. } \label{fig:civbc}
\end{figure}

\clearpage

\begin{figure}\figurenum{1}
\plotone{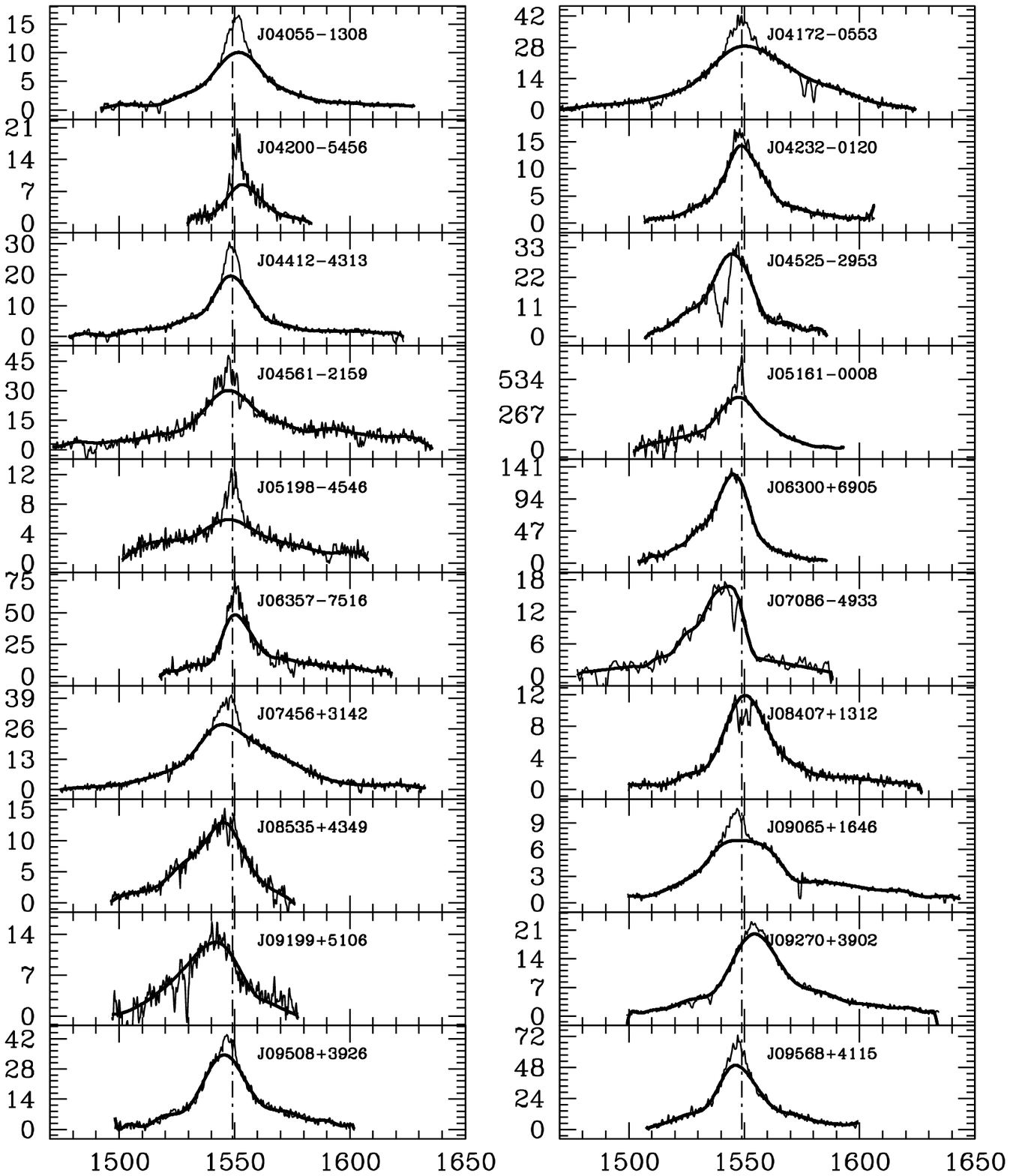} \caption{Cont.}
\end{figure}

\clearpage

\begin{figure}\figurenum{1}
\plotone{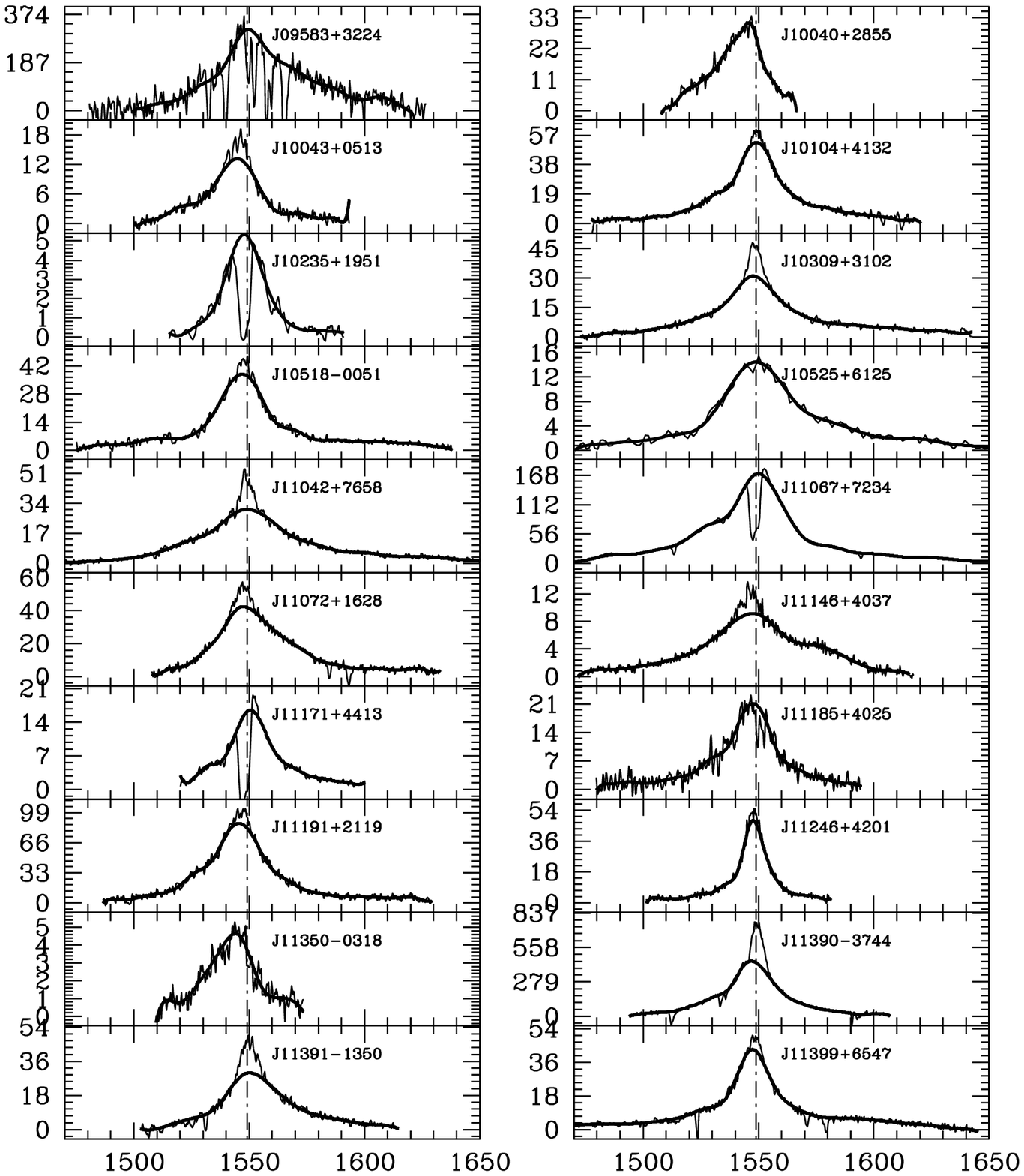} \caption{Cont.}
\end{figure}

\clearpage

\begin{figure}\figurenum{1}
\plotone{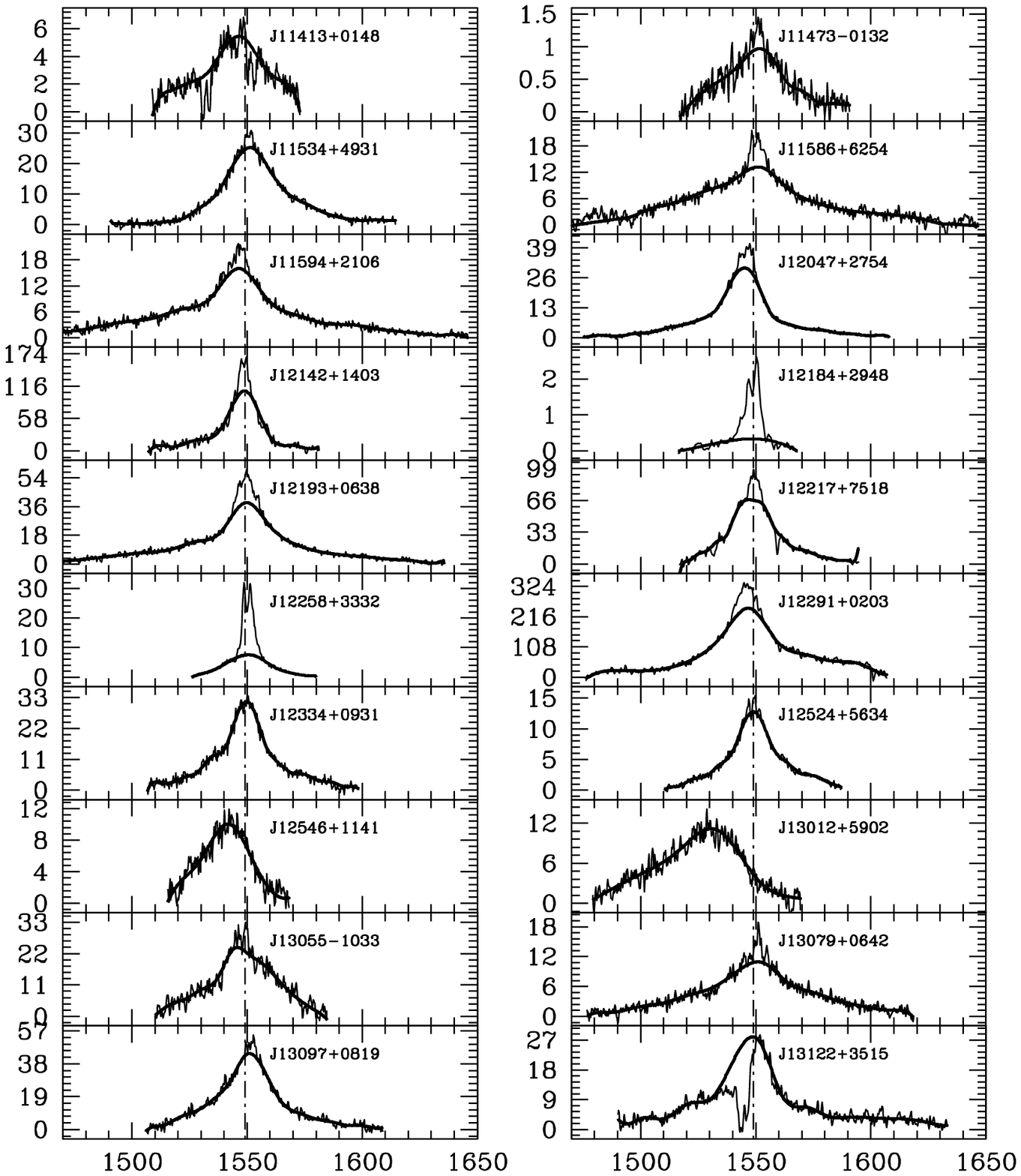} \caption{Cont.}
\end{figure}

\clearpage

\begin{figure}\figurenum{1}
\plotone{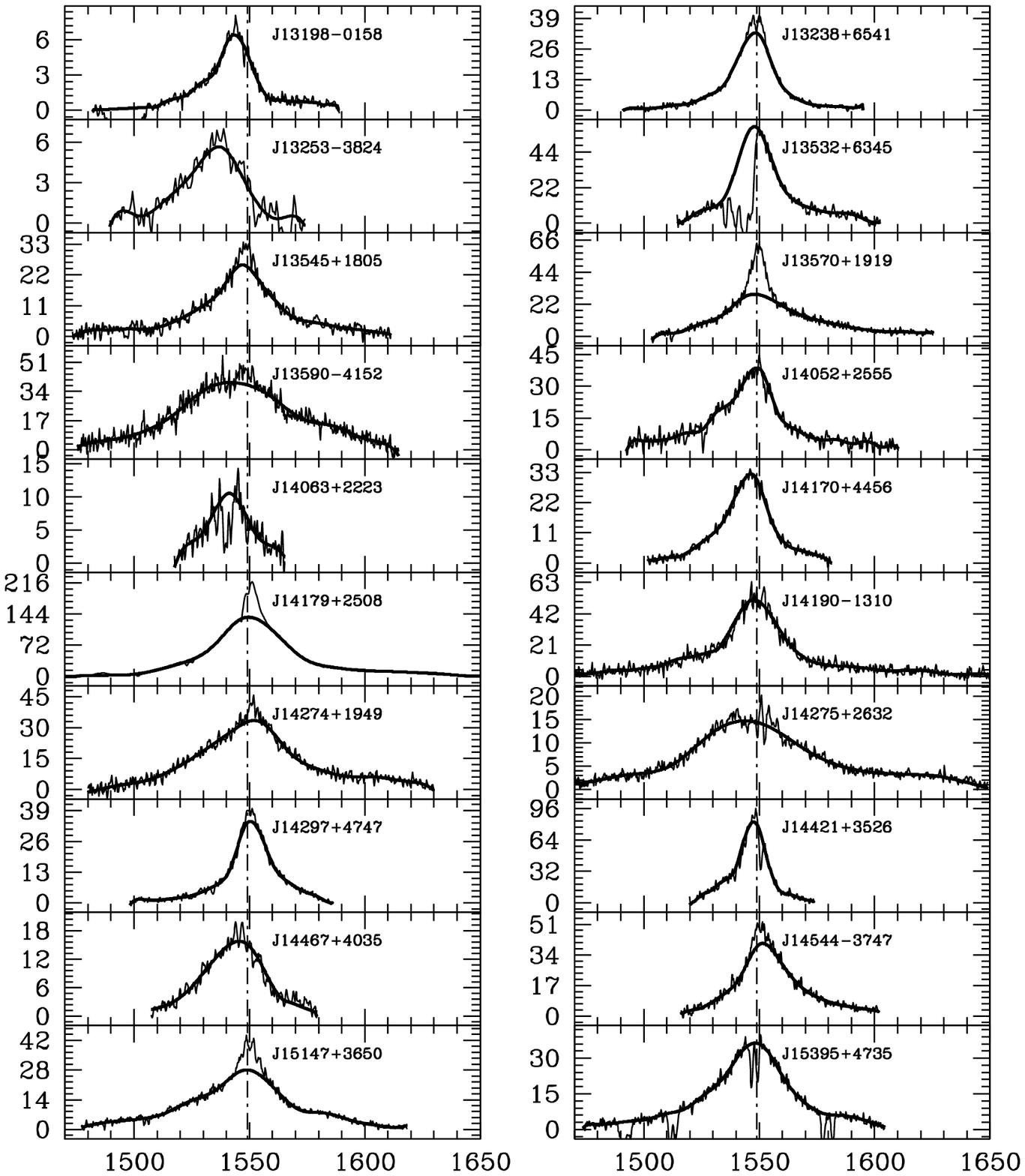} \caption{Cont.}
\end{figure}

\clearpage

\begin{figure}\figurenum{1}
\plotone{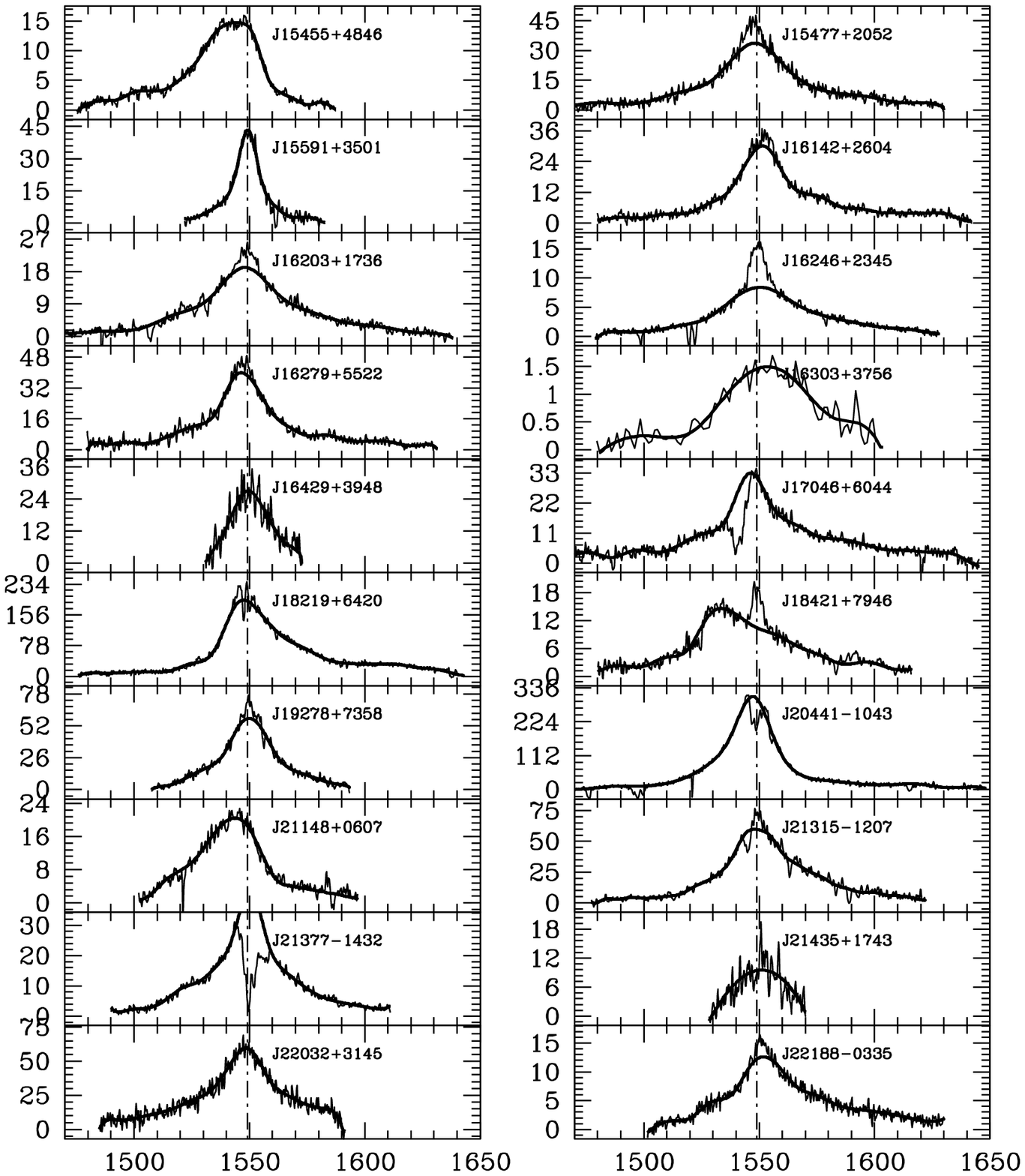} \caption{Cont.}
\end{figure}

\clearpage

\begin{figure}\figurenum{1}
\plotone{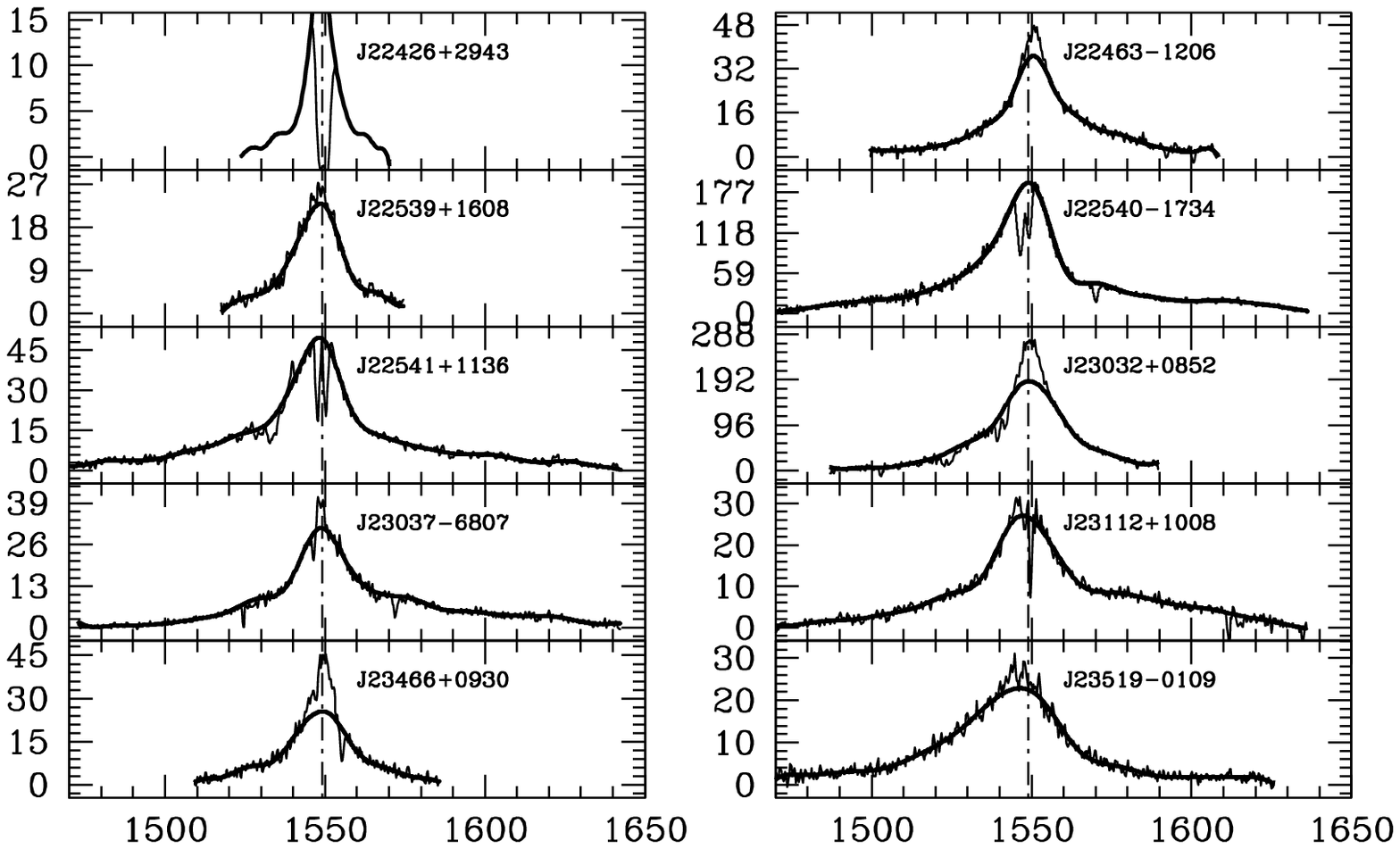} \caption{Cont.}
\end{figure}

\clearpage

\begin{figure}
\epsscale{0.4} \plotone{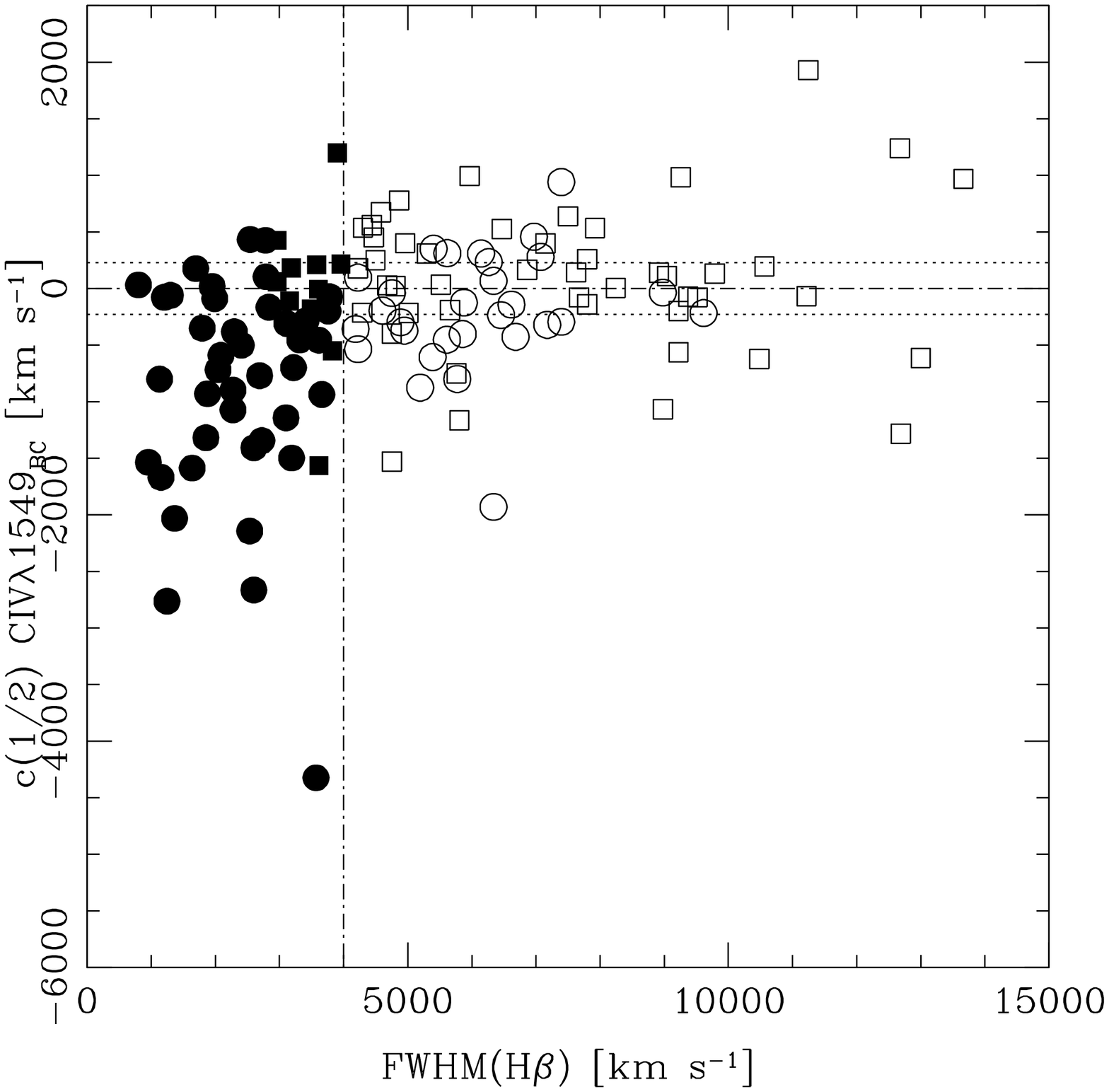} \epsscale{0.4} \plotone{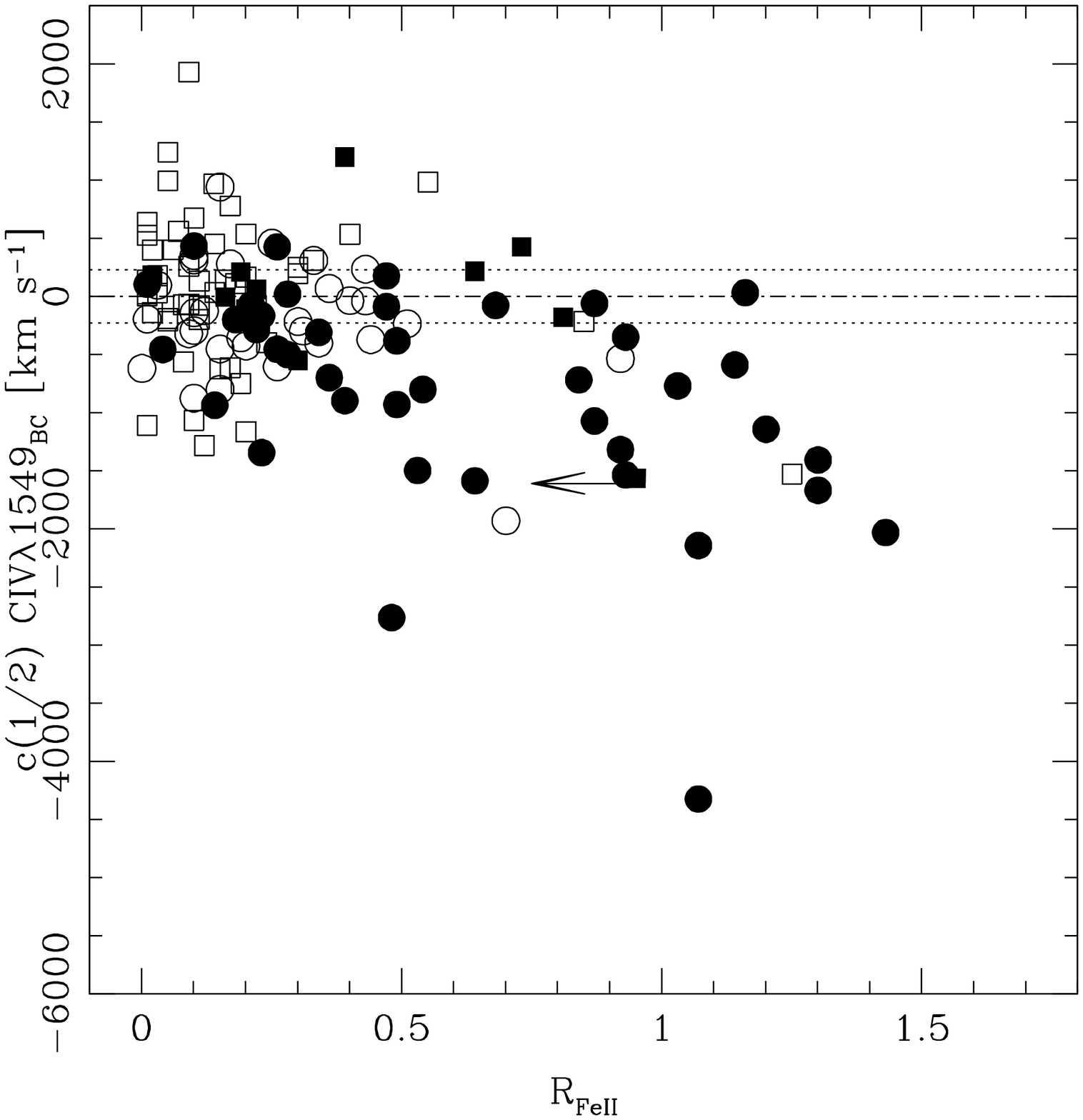}
\epsscale{0.4} \plotone{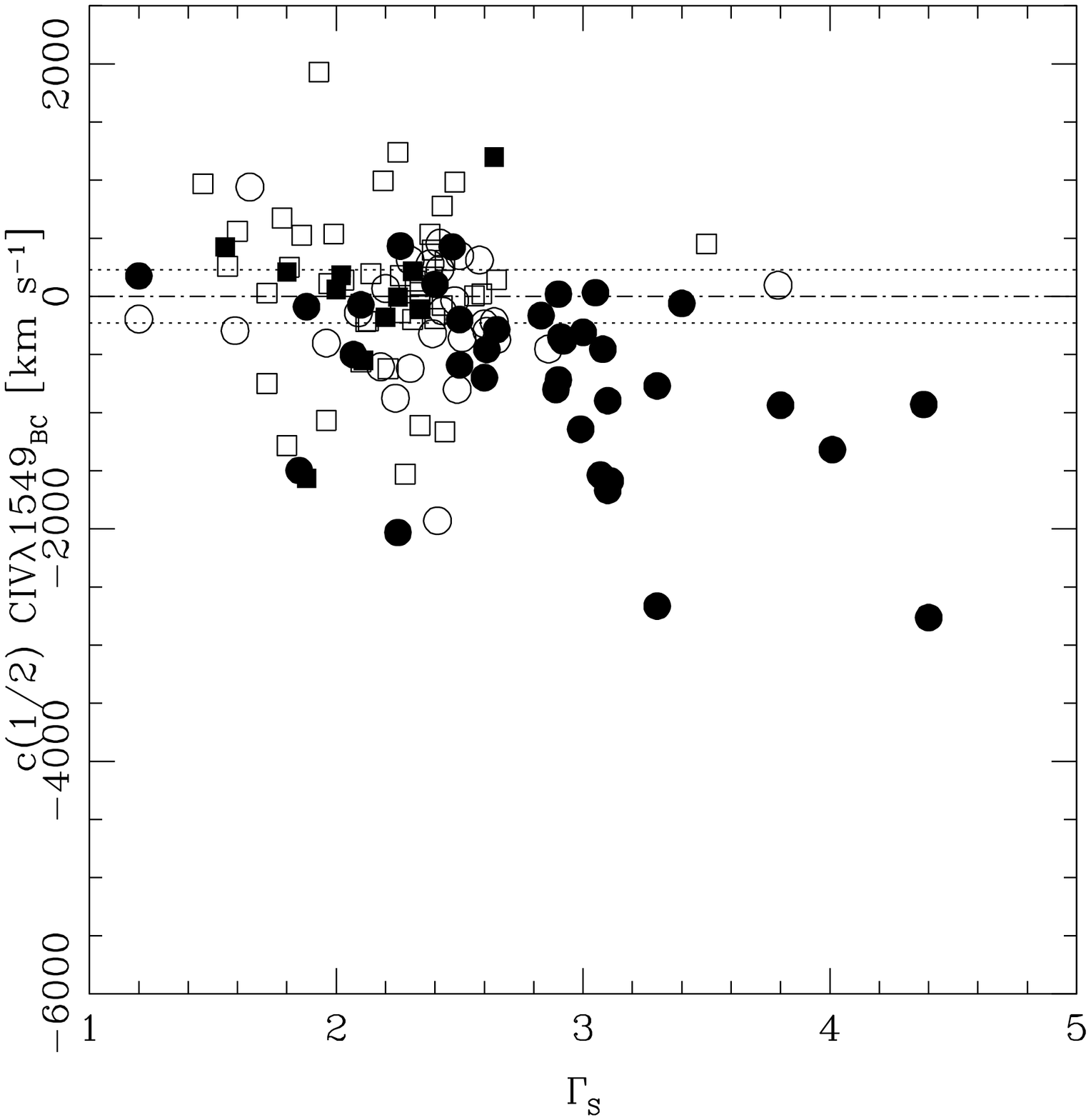} \epsscale{0.4} \plotone{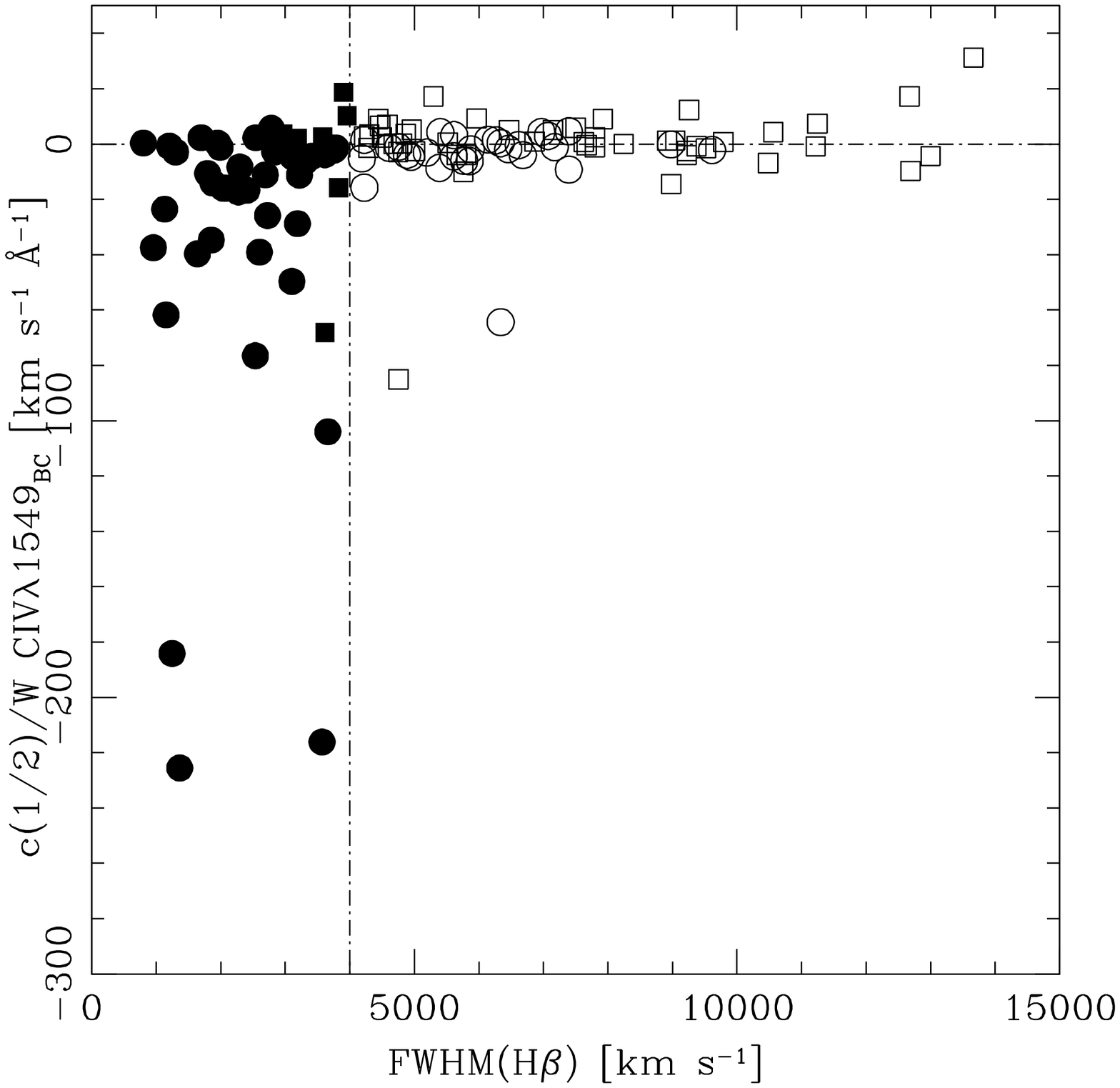}
\caption{4DE1 parameter planes involving \civbc\ profile shift at
half-maximum ($c(\case{1}{2})$, see text) vs. FWHM(\hbbc) (in
\kms) (UL), \rfe\ (UR) and  \gs\ (LL). In LR we also show
$c(\case{1}{2}$) normalized by EW \civbc\  in order to emphasize
the difference between Pop.  A and B sources which are denoted
with filled and open symbols respectively; radio-loud sources are
represented by squares and radio-quiet by circles. The vertical
line in the UL and LR panels marks the nominal Pop. A-B boundary.
Dotted lines indicate $\pm$ 2 $\sigma$ \ confidence intervals for
$c(\case{1}{2}$) (see \S \ref{immres}) meaning that sources within
that range do not show significant \civ\ line shift.}
\label{fig:e1}
\end{figure}

\clearpage

\begin{figure}
\epsscale{0.5} \plotone{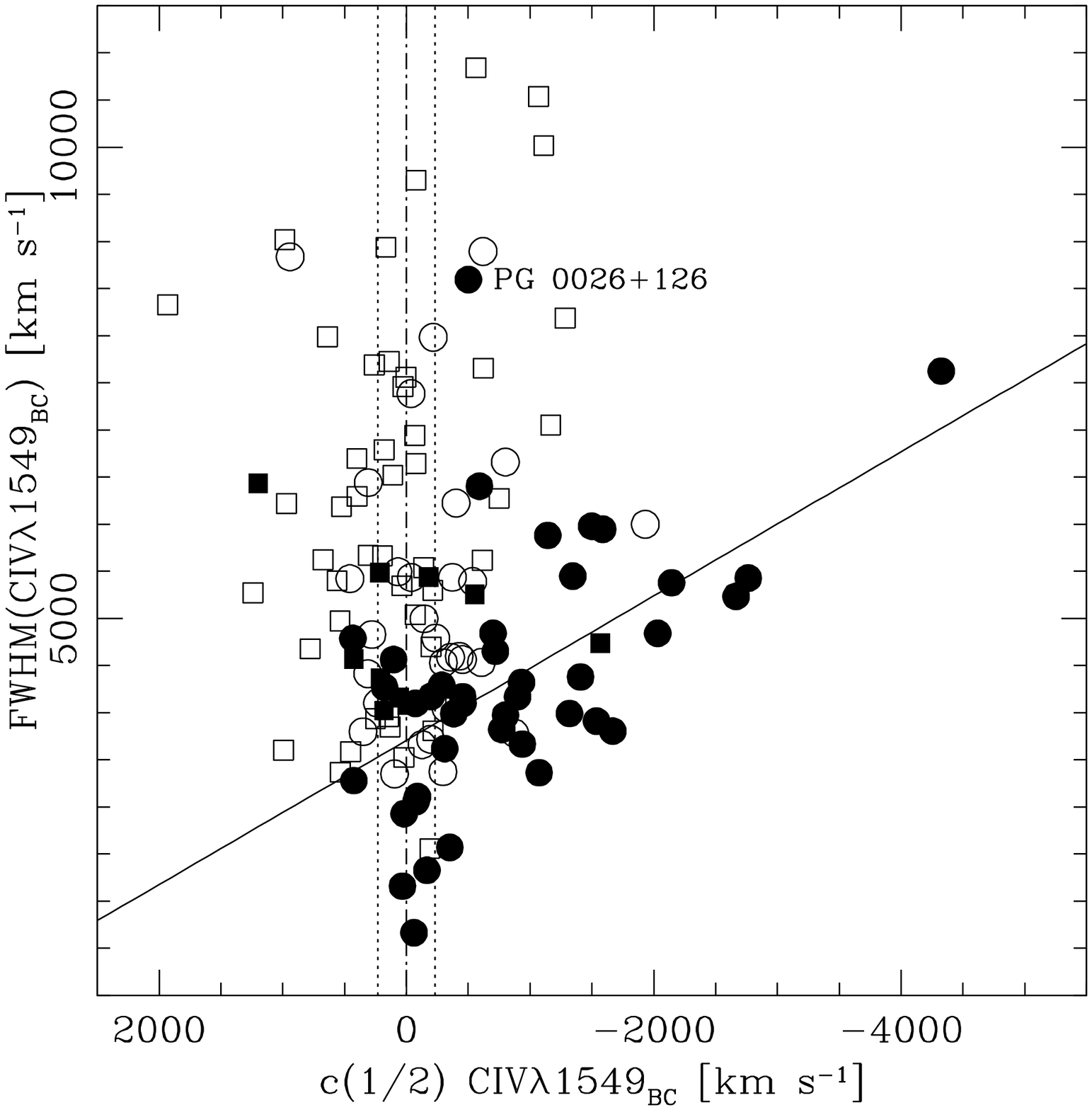} \plotone{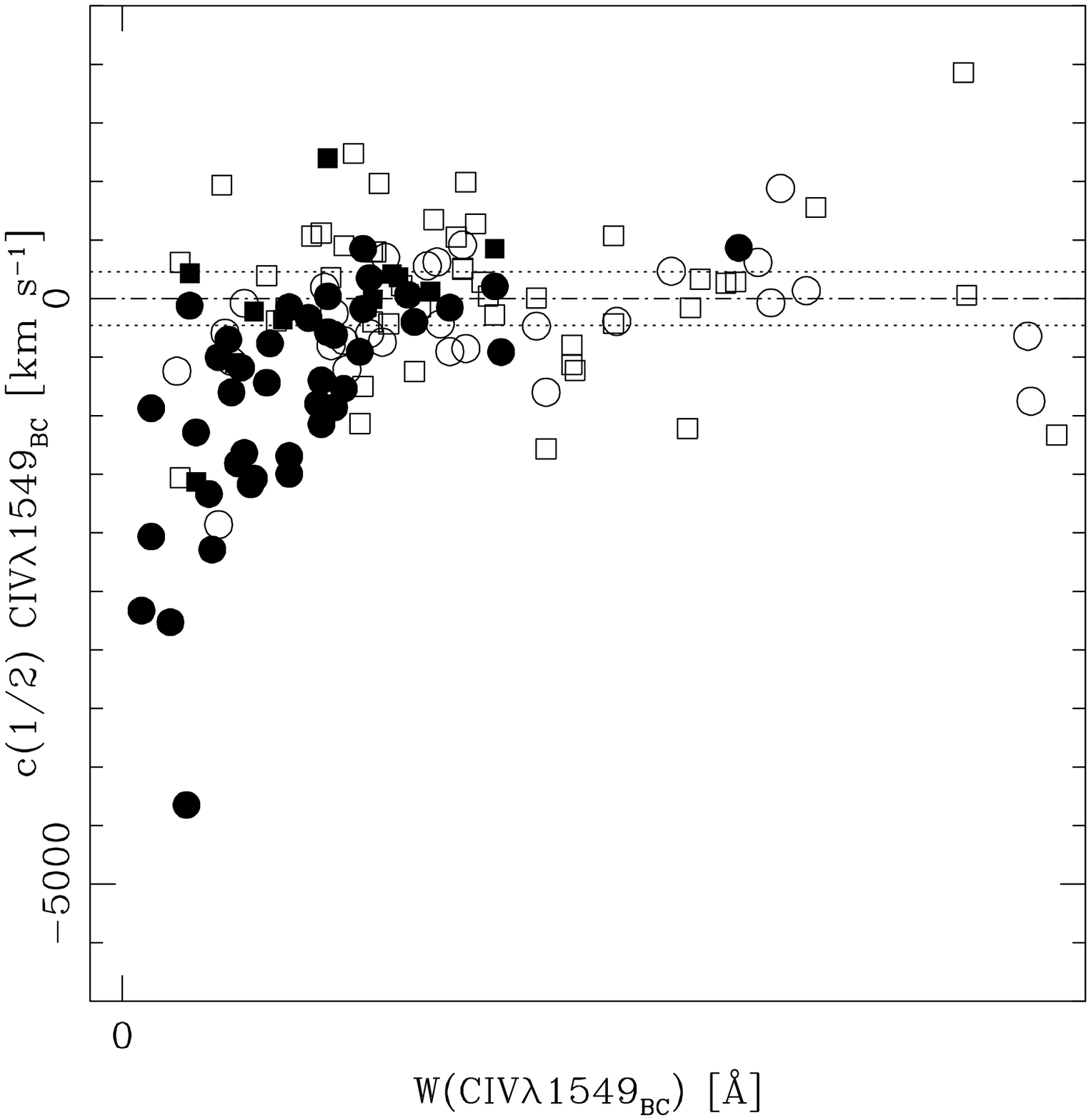} \caption{Upper
panel: Correlation diagram for measures of FWHM \civbc\ vs.
$c(\case{1}{2}$) for Pop. B  and Pop. A  sources. Symbols and
$c(\case{1}{2}$) confidence intervals are same as in the previous
Figure. The best fit regression line (lsq, unweighted) for the
Pop. A correlation (RQ only) is shown.  Both $c(\case{1}{2})$\ and
FWHM(\civbc) are in units of \kms. See Appendix \ref{appendix} for
a discussion of the outlier PG 0026+126. Lower panel:
$c(\case{1}{2})$\ vs. rest-frame W(\civbc). }\label{fig:civ}
\end{figure} \vfill

\clearpage

\begin{figure}
\epsscale{0.45} \plotone{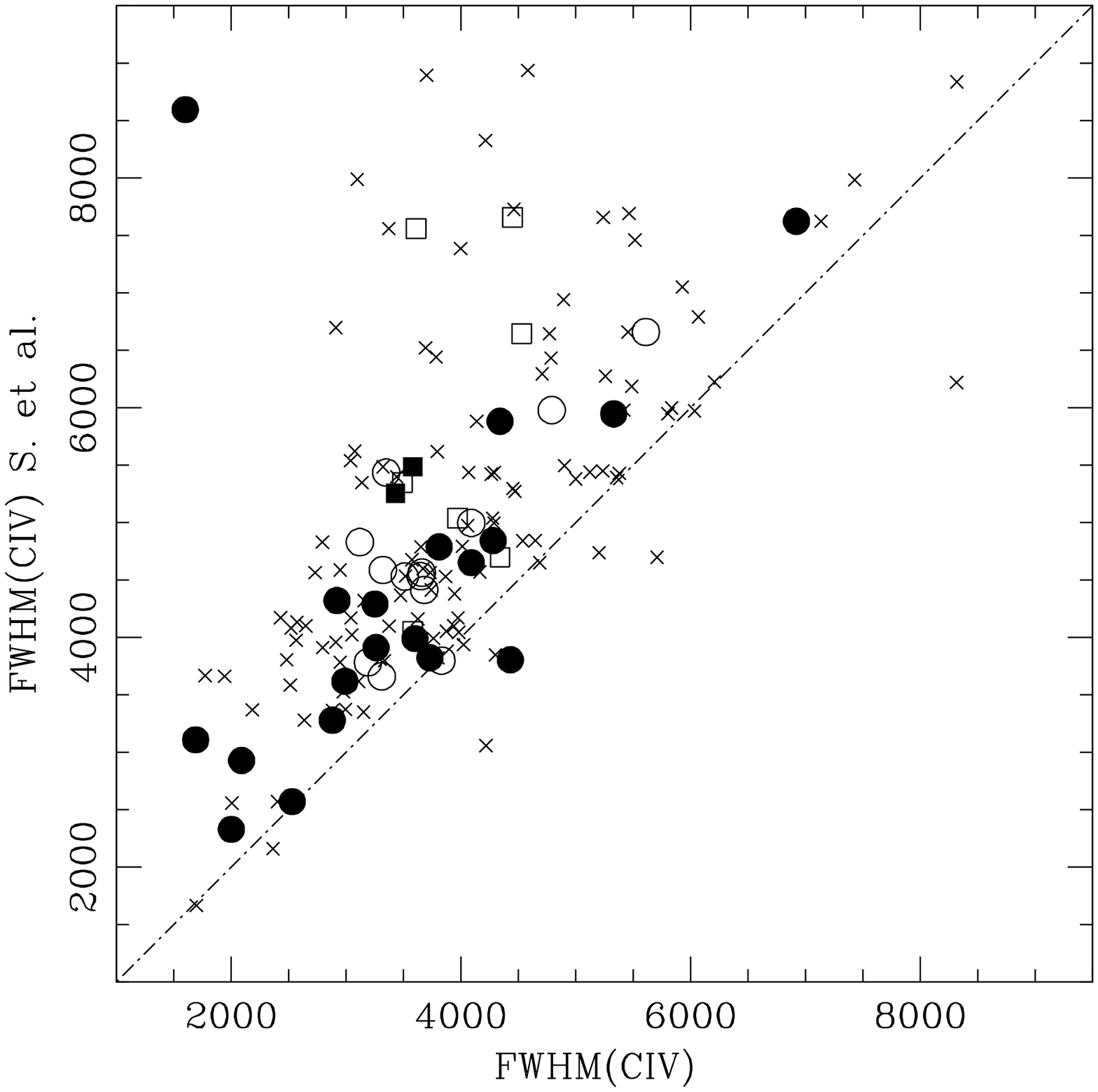} \plotone{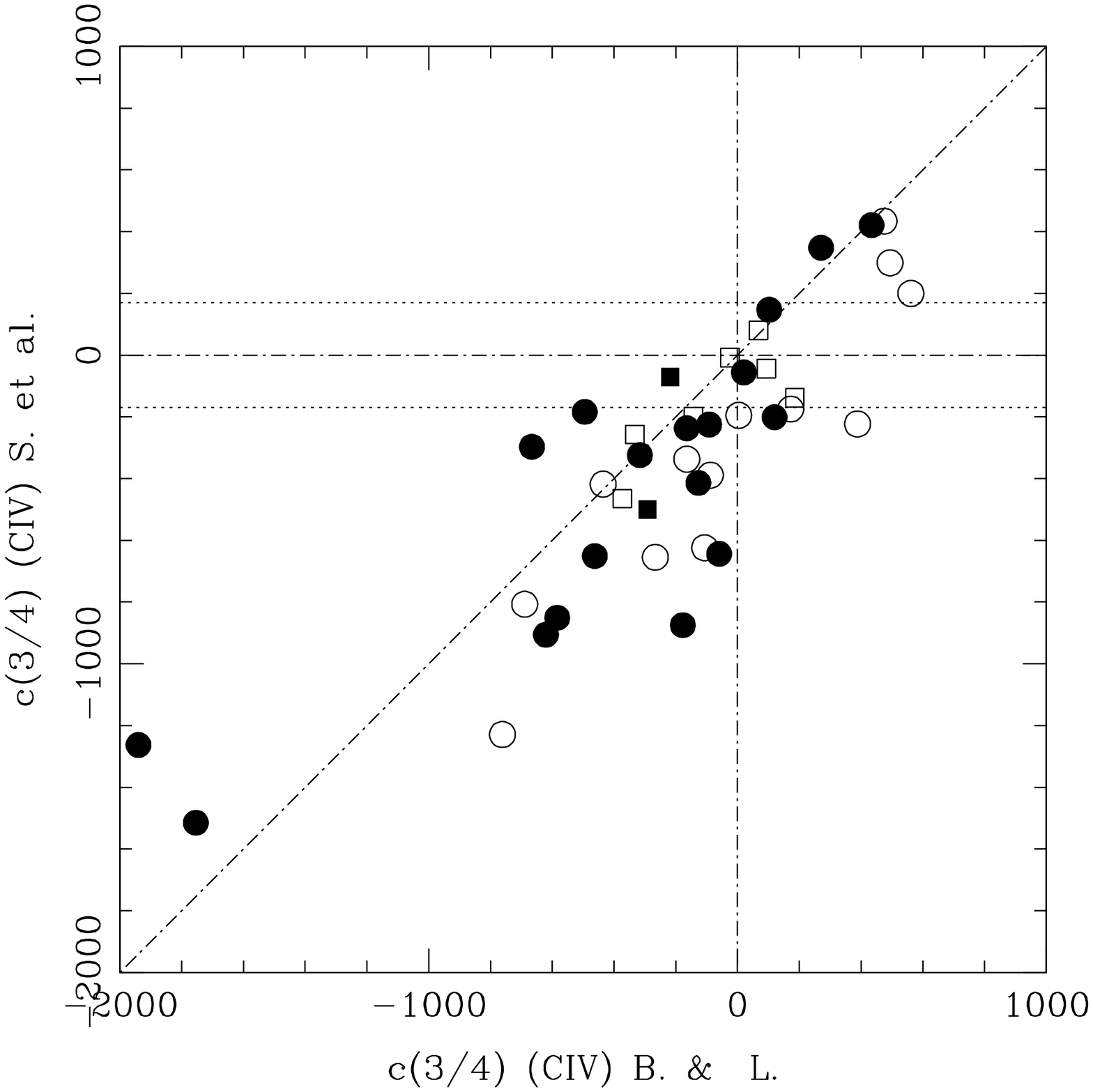}
\plotone{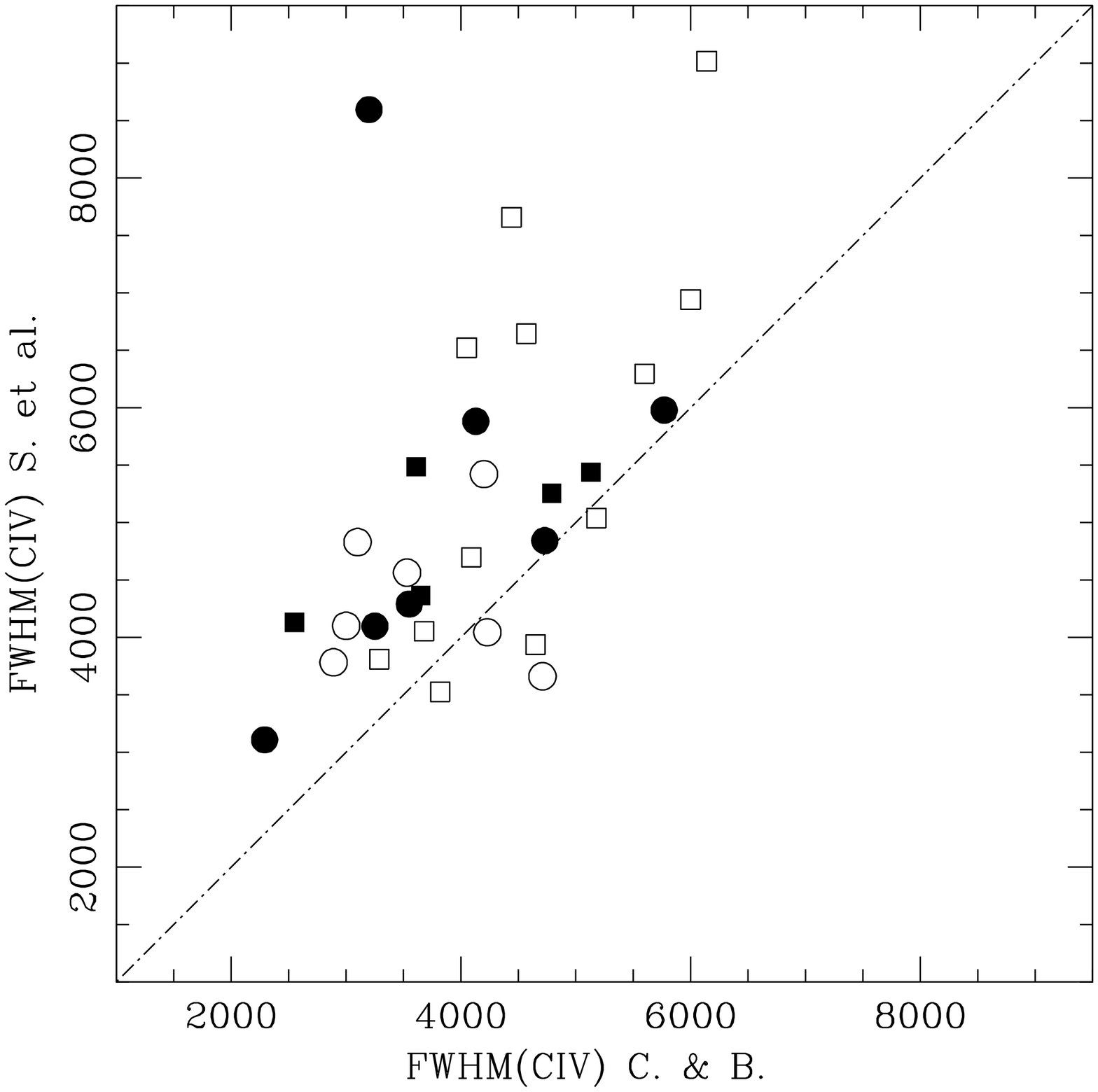} \plotone{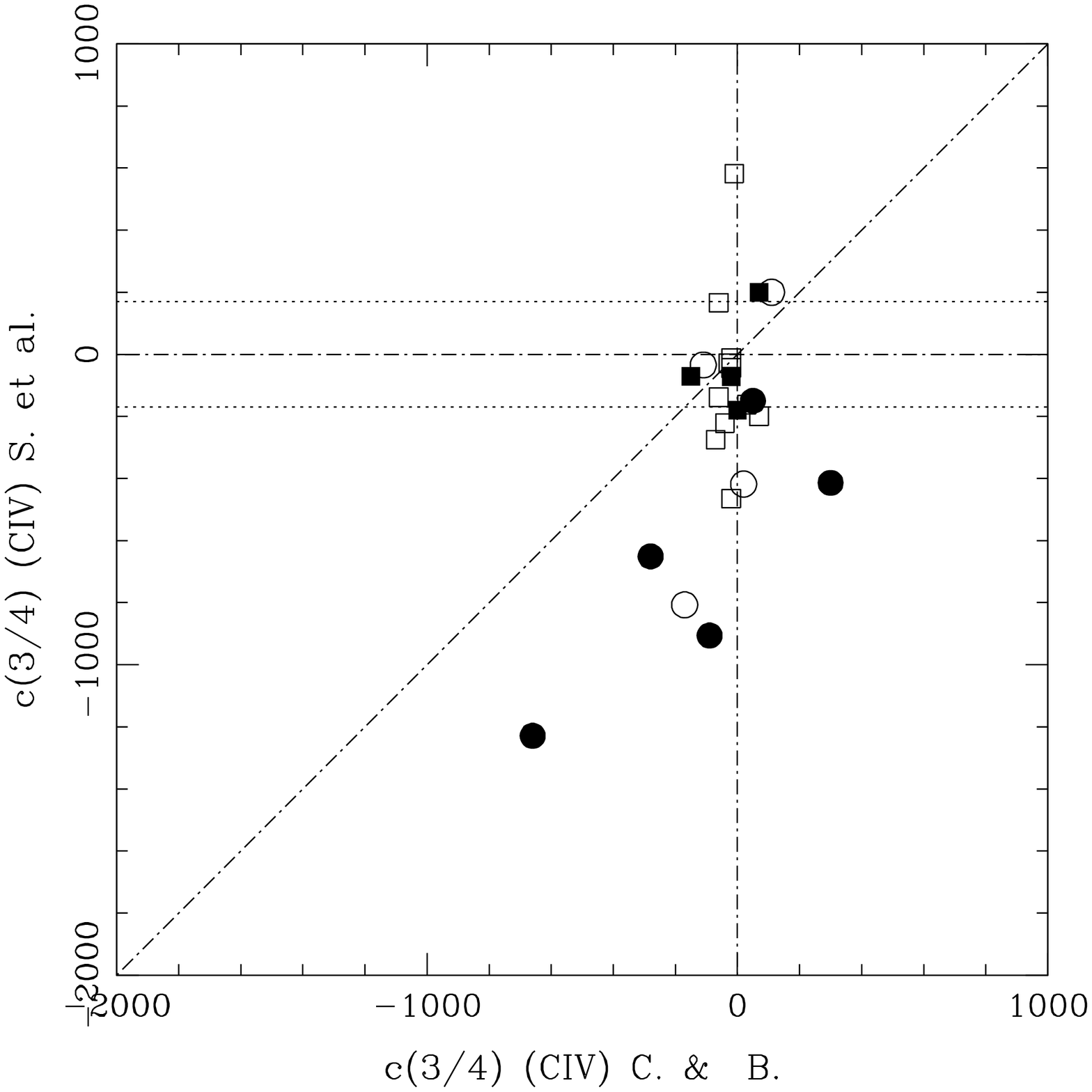} \caption{Comparison of our
\civbc\ measures with those of \citet{baskinlaor05} and
\citet{warneretal04} (upper panels) and \citet{corbor96} (lower
panels). Left panels: FWHM(\civbc) comparison, in units of \kms.
Small crosses compare with  FWHM(\civ) data of
\citet{warneretal04}. The same symbols used in previous figure
were used for comparisons with \citet{baskinlaor05}  and
\citet{corbor96}. Right panels: comparison \civ\ line centroid at
3/4 intensity ($c(\case{3}{4}$)), in \kms.  Our $c(\case{3}{4}$)
confidence intervals are shown in the right panels. The source
with the largest blueshift (Pop. A quasar PG 1259+593) falls
outside the boundary of the plot ($c(\case{1}{2}) \approx 4000$
\kms). Data point for PG 0026+129 is not shown to avoid x-scale
compression. Parity diagonal line is shown in all panels. Dotted
lines indicate $\pm$ 2 $\sigma$ \ confidence intervals for
$c(\case{3}{4}$) (see \S \ref{immres}).} \label{fig:bls}
\end{figure}

\clearpage

\begin{figure}
\epsscale{0.35}
\plotone{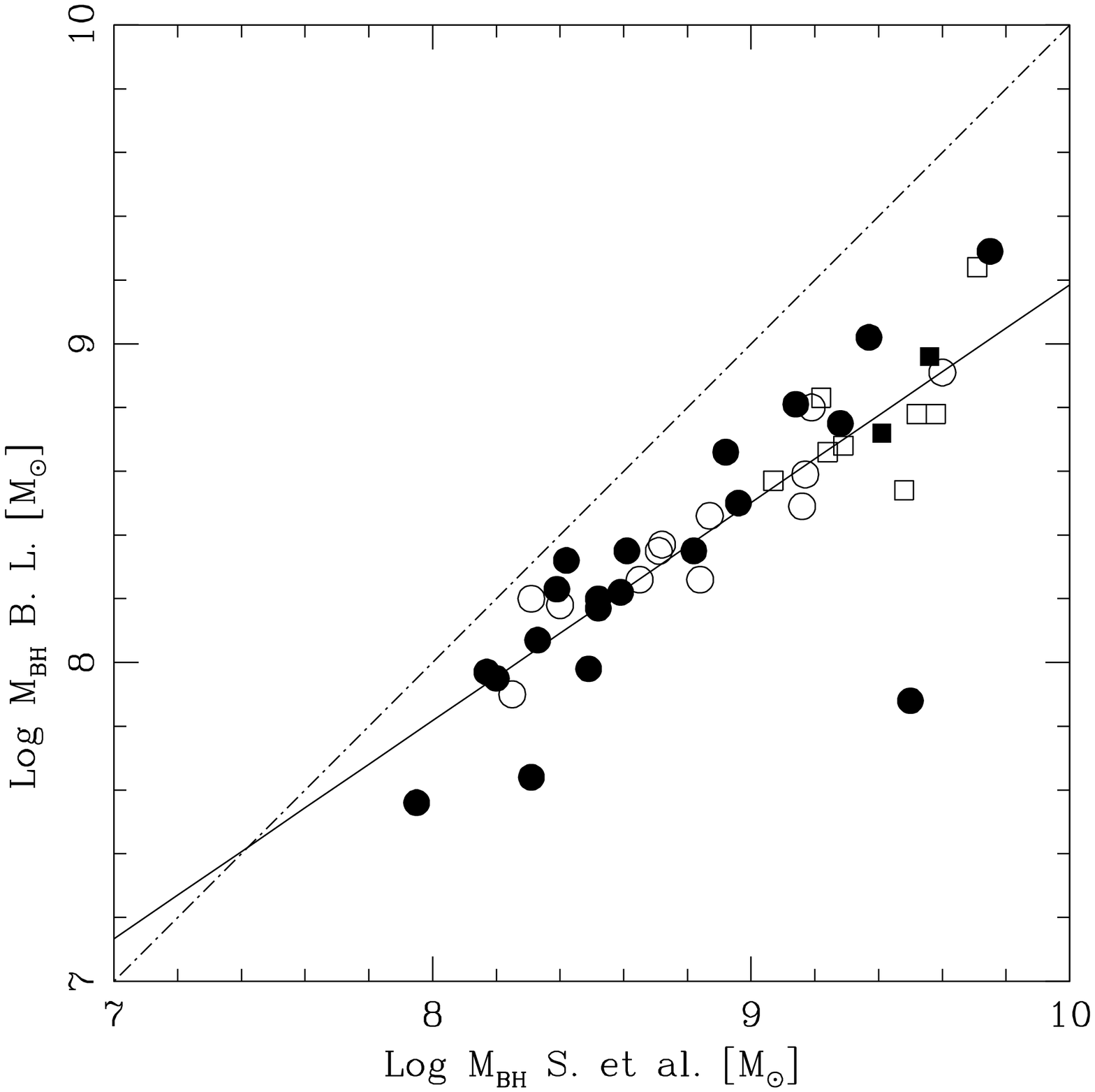}\\
\plotone{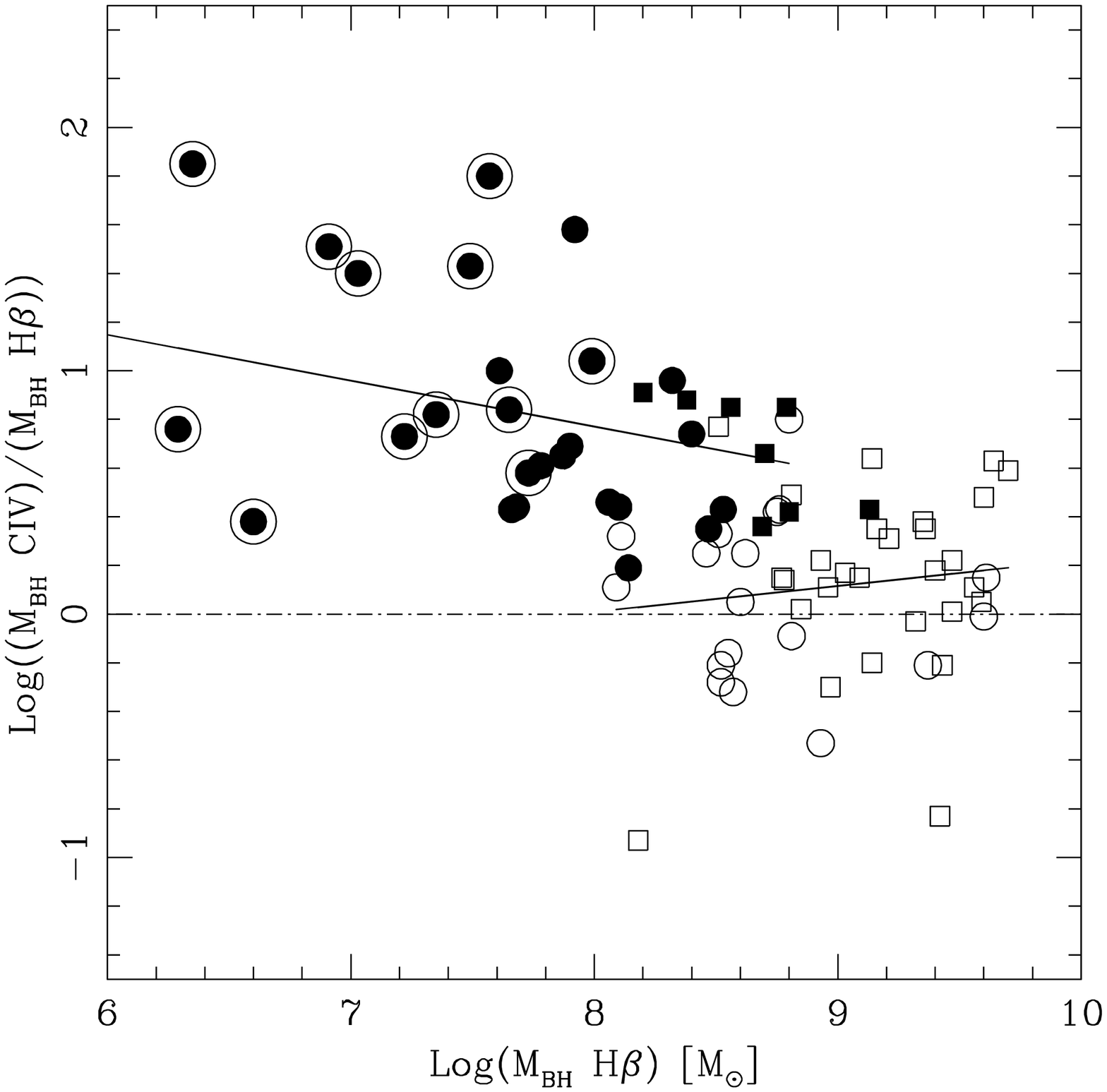}\\
\plotone{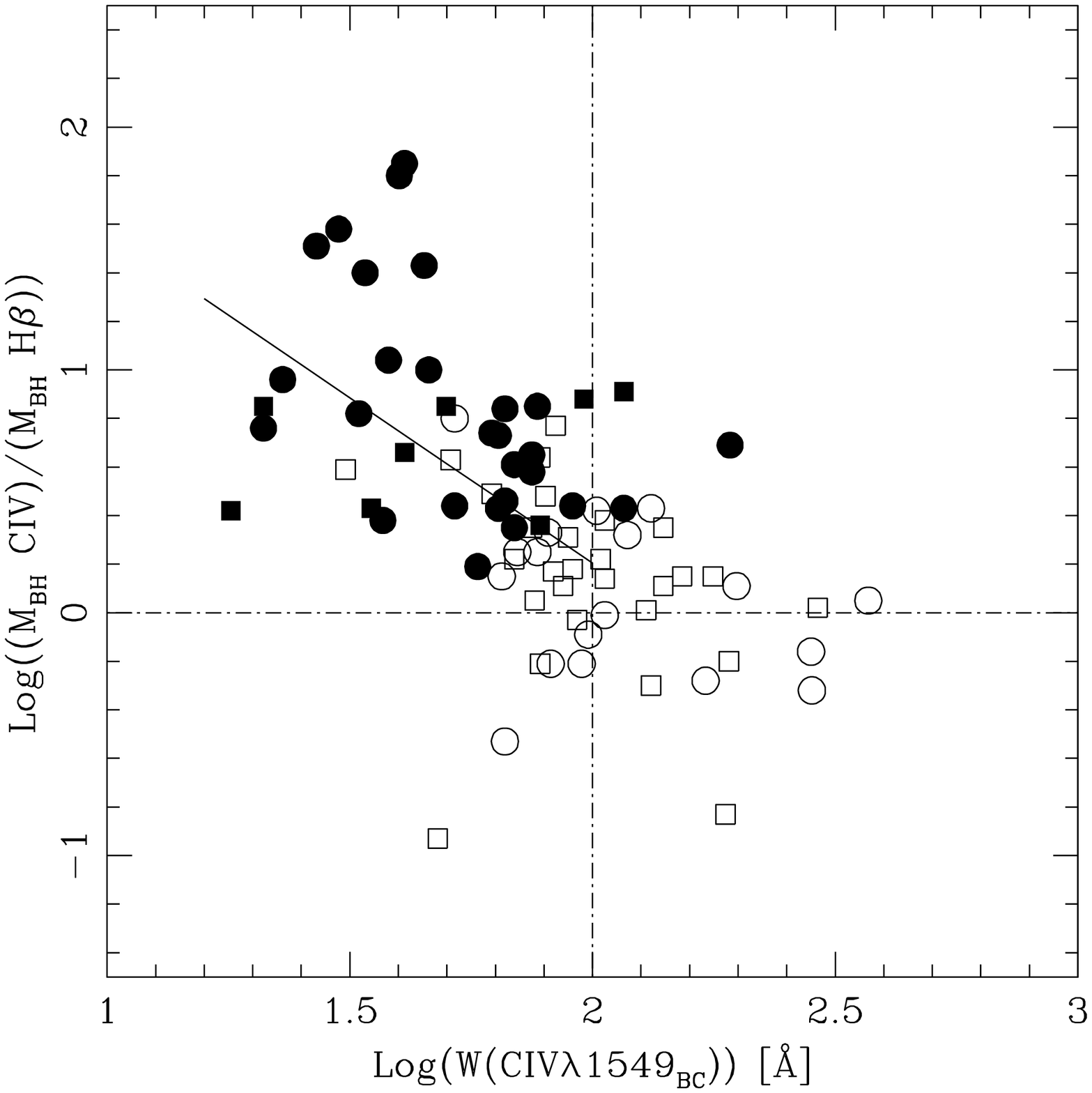} \caption{Upper panel: Comparison of our log
\mbh\ estimates for sources in common with \citet{baskinlaor05}.
``Virial'' velocity values derived from FWHM \civbc\ measures in
both samples. Dot-dash indicates parity line. The thin line shows
an unweighted lsq best fit for all sources. Middle panel: log
ratio of \mbh\ estimated from \civ\ and ``corrected" \hbbc\ (see
text in \S \ref{mbh}) vs. log \mbh for sources in common with
\citet{marzianietal03a}. Thin lines show independent best fits
(unweighted lsq) for Pop. A and B sources. NLSy1 sources are
identified among Pop. A sources by a larger open circle. Lower
panel: log ratio of \mbh\ estimated from \civ\ and ``corrected"
\hb\ as in panel above versus $\log W$(\civbc). Thin line shows a
best fit (unweighted lsq) for all sources with $\log W$(\civbc)$<$
2.} \label{fig:mbh}
\end{figure}

\clearpage

\begin{figure}
\epsscale{0.35} \plotone{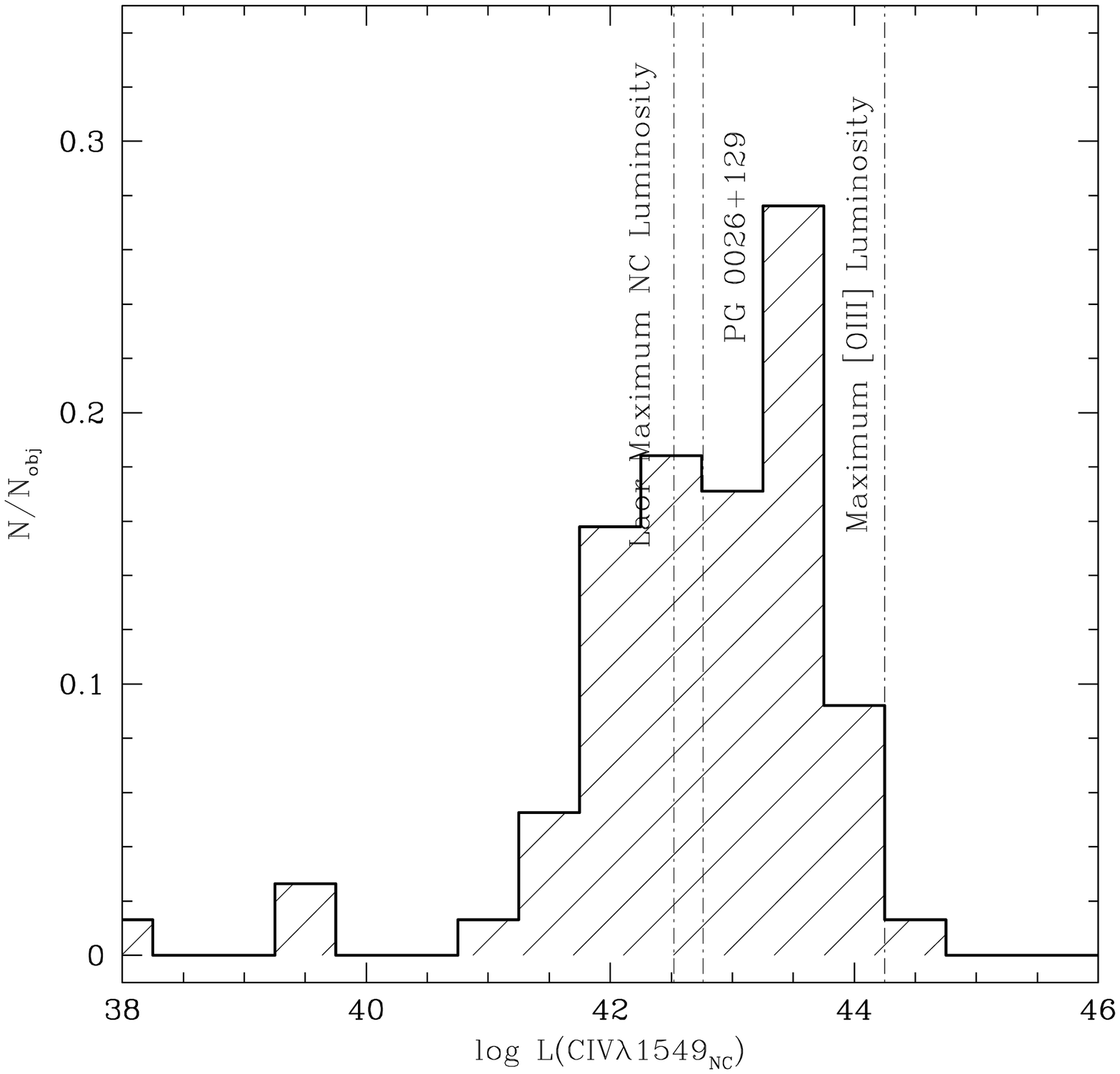} \epsscale{0.35}
\plotone{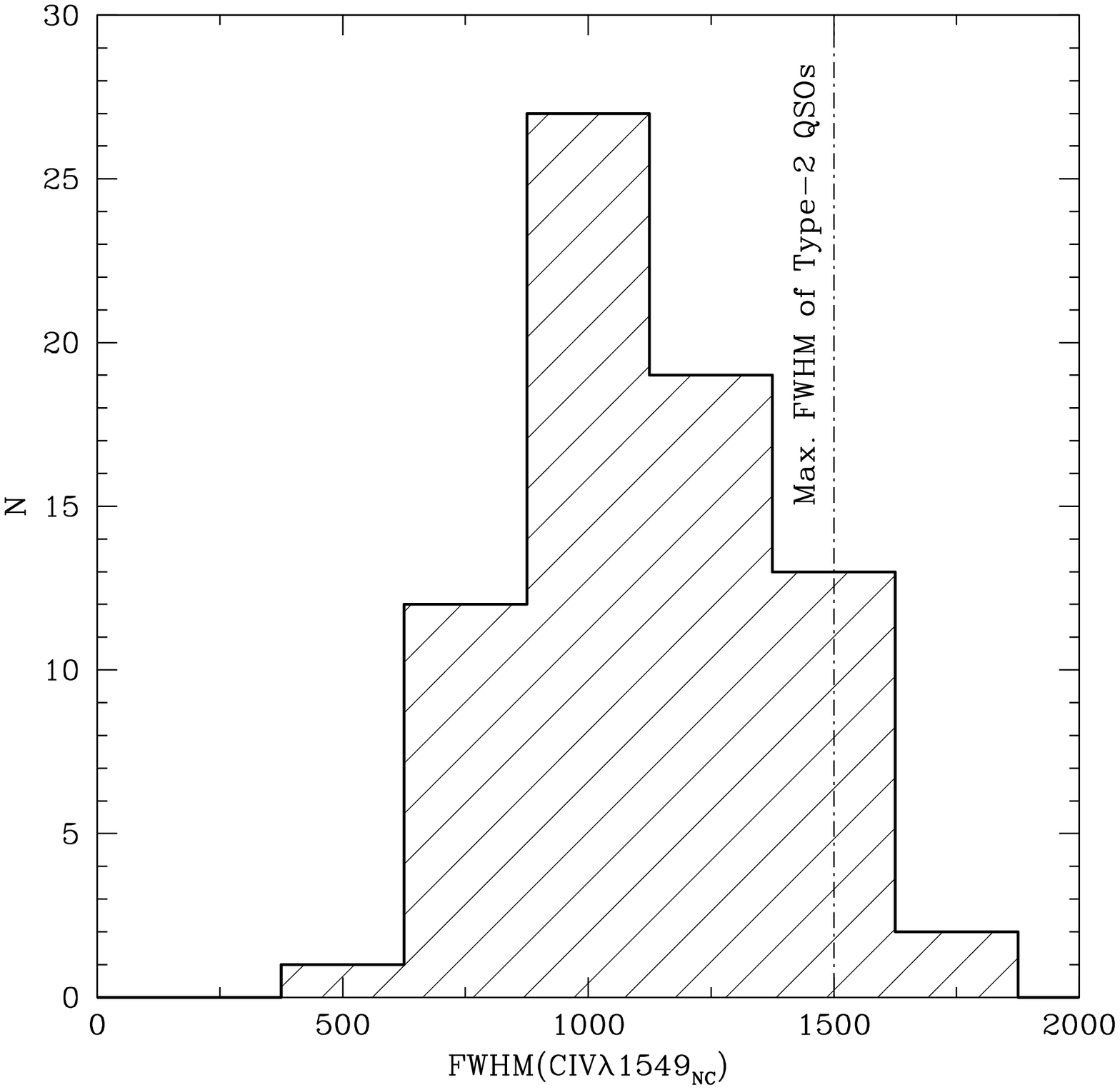} \epsscale{0.35}  \plotone{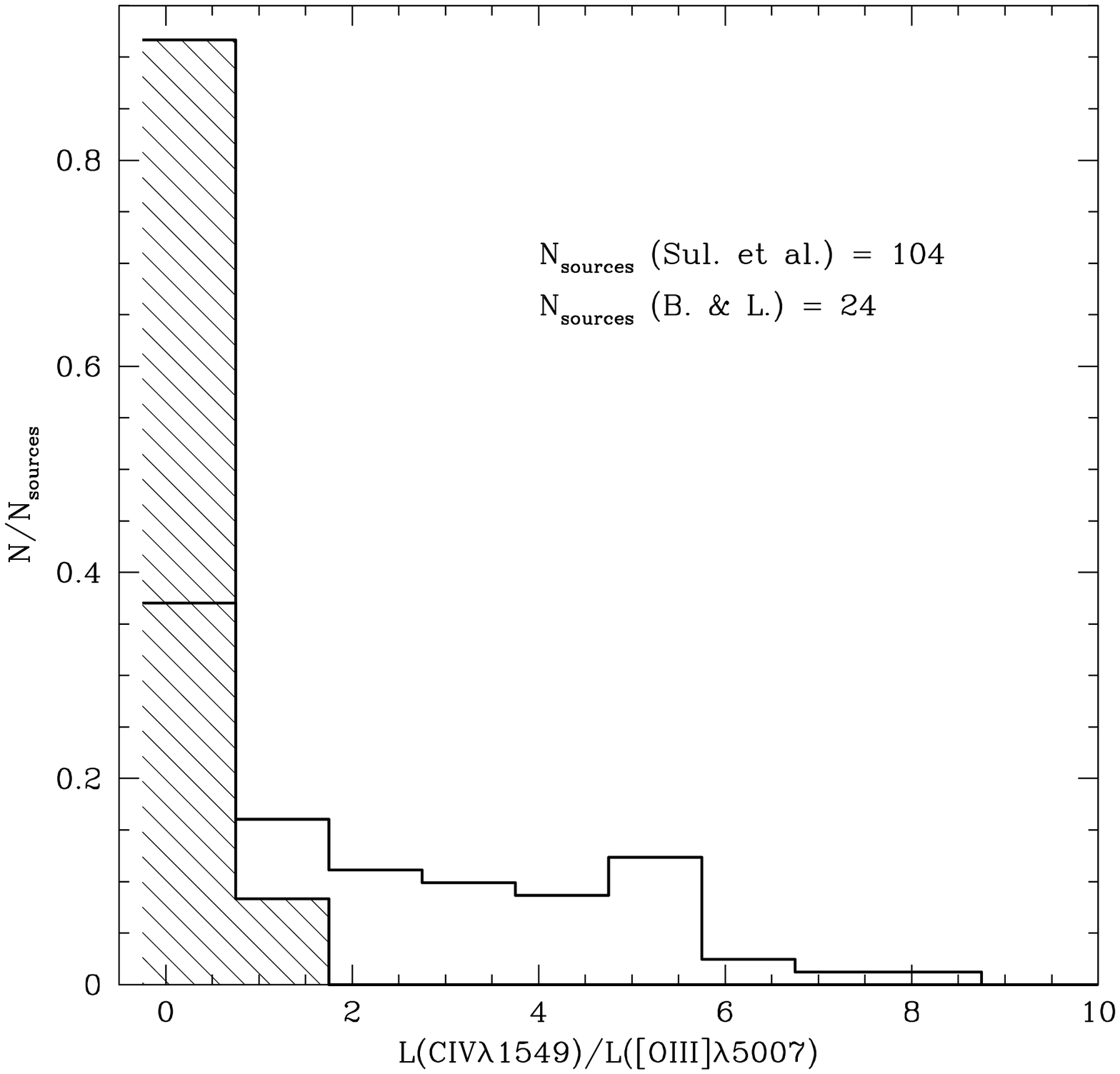}
\epsscale{0.35} \plotone{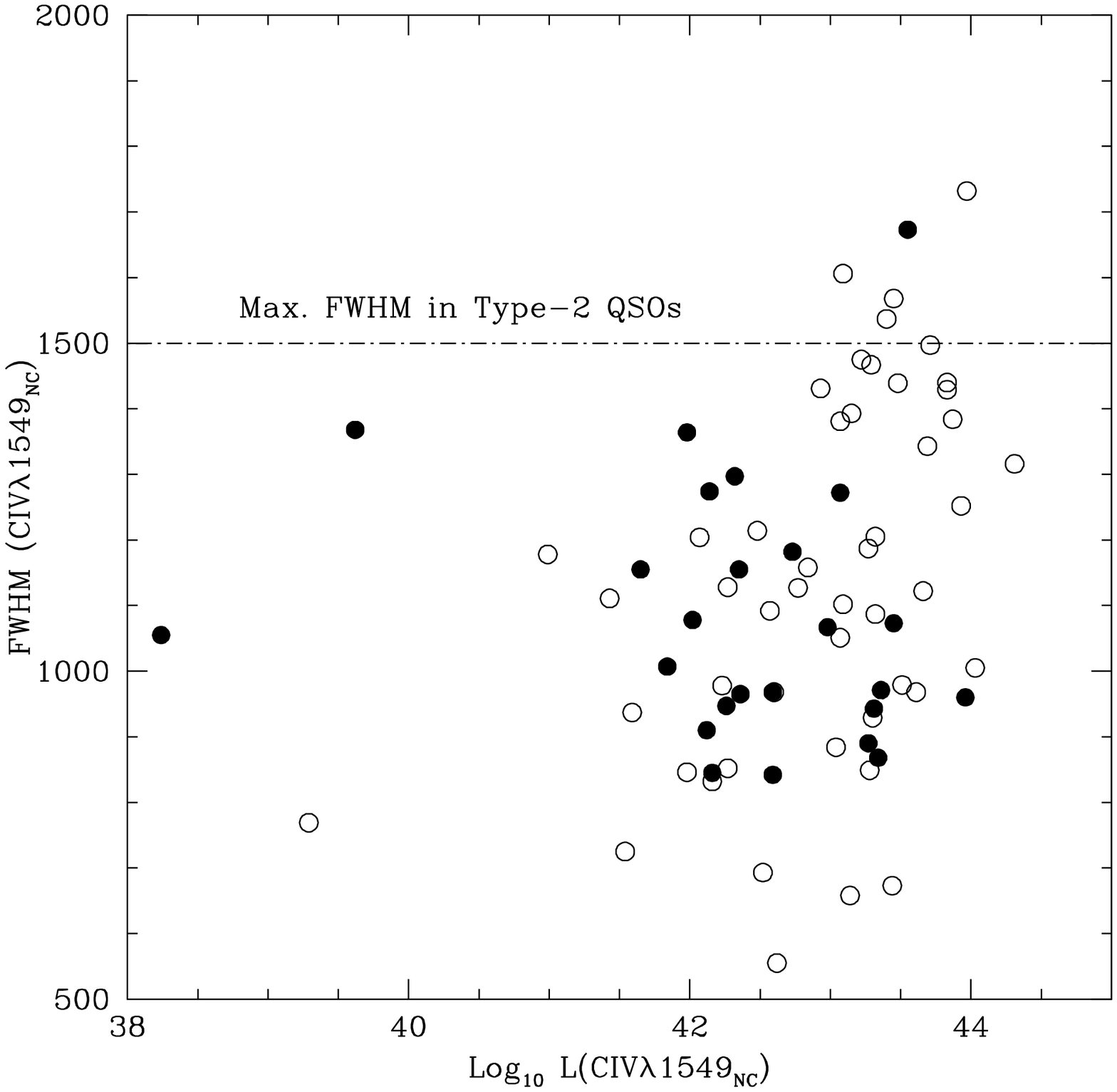} \caption{\civnc\ analysis. UL:
luminosity distribution of \civnc\ components identified in our
HST sample ($\log L$(\civnc) in units of \ergss; shaded
histogram); UR: FWHM distribution for \civnc\ components; LR:
distribution of the ratio L(\civnc)/L(\oiii) for our HST sample
(corrected for Galactic extinction) and for sample of
\citet{baskinlaor05}. LR: FWHM(\civnc) vs. $\log L$(\civnc) for
our HST sample. Filled circles indicate Pop. A, open circles Pop.
B sources. } \label{fig:civnc} \end{figure}

\end{document}